\documentclass[a4paper, 12pt]{article}
\usepackage{amssymb,
            amsmath,
            amsfonts,
            eurosym,
            geometry,
            ulem,
            color,
            setspace,
            comment,
            footmisc,
            pdflscape,
            subfigure,
            array,
            hyperref,
            xcolor,
            graphicx,
            bm,
            dcolumn}

\graphicspath{{outputs/}}                       
\usepackage[skip=3pt]{caption}

\usepackage{float}
\restylefloat{table}

\usepackage[affil-it]{authblk}
\usepackage[toc,page]{appendix}
\usepackage{mathtools}
\usepackage[round]{natbib}
\usepackage{mathrsfs}   

\usepackage{accents} 
\newlength{\dhatheight}

\newcommand\T{\rule{0pt}{2.6ex}}
\newcommand\B{\rule[-1.2ex]{0pt}{0pt}}

\usepackage{tikz} 
\usepackage{pgfplots}
  \usetikzlibrary{pgfplots.dateplot}
\usepackage{pgfplotstable}
  \pgfplotsset{compat=1.16}
  \usetikzlibrary{intersections,patterns,pgfplots.fillbetween}
\usetikzlibrary{decorations.pathreplacing} 
\usepackage{algorithm2e} 
\usepackage{amsthm}
\newcommand{\E}{\operatorname{\mathbb{E}}} 
\newcommand{\R}{\operatorname{\mathbb{R}}} 
\usepackage{bbm} 

\DeclareMathOperator*{\supp}{supp}
\DeclareMathOperator*{\p}{P}
\DeclareMathOperator*{\rank}{rank}
\DeclareMathOperator*{\st}{subject\,to}

\DeclareMathOperator*{\sign}{sign}

\DeclareMathOperator*{\var}{Var}

\DeclareMathAlphabet{\pazocal}{OMS}{zplm}{m}{n}

\newtheorem{theorem}{Theorem}
\newtheorem{corollary}{Corollary}
\newtheorem{assump}{Assumption}
\newtheorem{defin}{Definition}
\newtheorem{lem}{Lemma}
\newtheorem{prop}{Proposition}

\newcolumntype{L}[1]{>{\raggedright\let\newline\\arraybackslash\hspace{0pt}}m{#1}}
\newcolumntype{C}[1]{>{\centering\let\newline\\arraybackslash\hspace{0pt}}m{#1}}
\newcolumntype{R}[1]{>{\raggedleft\let\newline\\arraybackslash\hspace{0pt}}m{#1}}

\begin{document}

\begin{titlepage}
\title{Moran's $I$ Lasso for models with spatially correlated data \thanks{We are thankful to Hans-Martin Krolzig, Abhimanyu Gupta, Jorge Mateu, Pedro CL Souza, Nicolas Debarsy and Maria Kyriacou for discussions and suggestions. We would like to thank participants of the 20th International Workshop on Spatial Econometrics and Statistics, XV World Conference of Spatial Econometrics and Econometric Society - Delhi Winter School for valuable comments and discussions. All remaining error are the authors'.}}

\author{By Sylvain Barde, Rowan Cherodian and Guy Tchuente\thanks{Barde: University of Kent; Cherodian (corresponding author): Regent Court (ScHARR), University of Sheffield, S1 4DA, \texttt{r.cherodian@sheffield.ac.uk}; Tchuente: Purdue University}}

\date{\today}
\maketitle

\begin{abstract}
This paper proposes a Lasso-based estimator which uses information embedded in the Moran statistic to develop a selection procedure called Moran's $I$ Lasso (Mi-Lasso) to solve the Eigenvector Spatial Filtering (ESF) eigenvector selection problem. ESF uses a subset of eigenvectors from a spatial weights matrix to efficiently account for any omitted cross-sectional correlation terms in a classical linear regression framework, thus does not require the researcher to explicitly specify the spatial part of the underlying structural model. We derive performance bounds and show the necessary conditions for consistent eigenvector selection. The key advantages of the proposed estimator are that it is intuitive, theoretically grounded, and substantially faster than Lasso based on cross-validation or any proposed forward stepwise procedure. Our main simulation results show the proposed selection procedure performs well in finite samples. Compared to existing selection procedures, we find Mi-Lasso has one of the smallest biases and mean squared errors across a range of sample sizes and levels of spatial correlation. An application on house prices further demonstrates Mi-Lasso performs well compared to existing procedures. 
\end{abstract}


\bigskip


\bigskip

\setcounter{page}{0}
\thispagestyle{empty}
\end{titlepage}

\pagebreak \newpage

\doublespacing

\section{Introduction} \label{sec:introduction}

In conventional spatial economic modeling, the researcher is required to specify (i) a spatial weights matrix (SWM)\footnote{A spatial weights matrix is an $n\times n$ matrix  describing the pair-wise relationships between the $n$ cross-sectional units.} and (ii) which parts of the model are spatially correlated. Standard specifications typically include one or more spatial lags of the dependent, exogenous and/or error term \citep{kel17}. Historically, applied researchers have generally focused more on specifying the SWM rather than the empirical spatial structure \citep{lsK_sens14} and a standard robustness check in the applied spatial economic literature tests whether the estimates are sensitive to different SWMs. When estimates are found to be sensitive to the choice of SWM, researchers have attributed this sensitivity to the choice of SWM. However, as \citet{lsK_sens14} shows, estimates should not be overly sensitive to the choice of SWM, as long as they are reasonably well correlated. This implies that the sensitivity many researchers observe is driven by misspecification of the spatial economic model rather than the choice of SWM. \citet{lsK_sens14} thus argue that researchers should focus on specifying the spatial model rather than finding an ideal SWM.

The Eigenvector Spatial Filtering (ESF) approach of \citet{grif2000, grif03} uses a subset of eigenvectors from the SWM as controls to filter out terms involving the SWM in the underlying model. ESF has recently started receiving substantial attention from applied economic researchers.\footnote{Some examples include \citet{PGTN11_ESF, RSGN_JRS12ESF, esfJAE13, CS14ESF, ESF_sjpe18, bg_bcon08esf, gp11ESF}.} ESF's main advantage over conventional maximum likelihood (ML) and generalised method of moments (GMM) is precisely that researchers need not specify spatial correlation explicitly in the model, or estimate the corresponding spatial parameters. In the context of ESF these are instead viewed as nuisance parameters. The fact that ESF is agnostic to the underlying spatial process is desirable for applied researchers who simply wish to obtain unbiased parameter estimates in the presence of cross-sectional dependence in the covariates. This is because it is easier to establish the presence of an underlying spatial process via a test for spatial correlation than it is to determine the exact specification of this process.

The critical challenge for ESF is that the spectral decomposition of the $n\times n$ SWM yields $n$ eigenvectors and if all are included in the model, it becomes high-dimensional and estimation by Ordinary Least Squares (OLS) is infeasible.\footnote{A high-dimensional model is defined as a model with more parameters to estimate than observations, leading to a rank-deficient Gram matrix.} \citet{grif03} argues that only a subset of eigenvectors is necessary to eliminate the cross-sectional dependence in the dependent variable. The key question becomes identifying which subset of eigenvectors is required, which we refer to as the ESF eigenvector selection problem. Several solutions to this selection problem have been proposed, such as several stepwise greedy algorithms where eigenvectors are iteratively added until some user-specified threshold is reached \citep{grif2000,grif03,tiefgrif07}. These stepwise greedy algorithms are simply heuristic approximations to the full ESF selection problem, thus, they are necessarily sub-optimal. Under the assumption of sparsity (i.e. most eigenvector coefficients are zero) \citet{seya15} proposes using an $\ell_1$-penalised regression, e.g. Lasso. Given that Lasso estimates are ultimately determined by a tuning parameter, this turns the eigenvector selection problem into a tuning parameter calibration problem. \citep{seya15} propose estimating the tuning parameter using conventional $K$-fold cross-validation (CV) with prediction accuracy as the loss function. However, the existing theoretical results on CV-Lasso assume the cross-sectional units are independent \citep{czc20}. This is hard to justify in the context of ESF, where the eigenvectors are derived from a matrix that encodes cross-sectional dependence. Additionally, the goal of ESF is to eliminate spatial correlation patterns, not improve prediction accuracy. There is therefor no guarantee that running CV with a prediction accuracy loss will yield consistent eigenvector selection.

We propose an alternative procedure for choosing the ESF Lasso tuning parameter,  called Moran's $I$ Lasso (Mi-Lasso), which directly uses information about the level of correlation in the residuals provided by the Morans $I$ statistic \citep{moran50} to develop a point estimate for the Lasso tuning parameter. The intuition behind Mi-Lasso is that when the spatial correlation in the residuals is low, only a small set of eigenvectors will be necessary, so a high level of regularisation is required, and vice versa for a high level of residual spatial correlation. Mi-Lasso has several advantages; the method is (i) intuitive, (ii) theoretically grounded, and (iii) substantially faster than Lasso with $K$-fold cross-validation (CV) or the stepwise iterative greedy algorithms suggested in the literature.\footnote{Mi-Lasso only requires estimating a single point on a Lasso path, unlike $K$-fold cross validation which requires estimating $K$ paths. The larger $K$, the more computationally demanding the procedure.}

We establish the theoretical properties of Mi-Lasso by formalising the implicit ESF assumption that the terms which include the SWM can be approximated by a subset of eigenvectors. Under some standard spatial regularity conditions, we then derive non-asymptotic bounds for the coefficients of the eigenvectors and also assess the additional conditions required for Mi-Lasso to yield consistent eigenvector selection. In addition, given that the spectral decomposition of a square matrix power $\bm{A}^p$ always produces the same eigenvectors $\forall \; p \in \mathbb{Z}^+ $, we show that ESF also handles the case where the unknown spatial process possesses higher-order lags of the SWM. Simulations confirm that Mi-Lasso performs well for a range of levels of spatial correlation and when the data-generating process includes higher-order lags. Regarding computational time, Mi-Lasso is at least an order of magnitude faster than CV-Lasso.\footnote{Our setup explores sample sizes up to $n=10^4$, at which point the forward stepwise procedures become infeasible.}

Finally, we examine the practical performance of Mi-Lasso with an empirical application using the Boston Housing Dataset. We find that Mi-Lasso selects more than triple the number of eigenvectors compared to existing procedures. However, Mi-Lasso gives a bitter fit of the data in terms of adjusted $R^2$ and has substantially fewer insignificant eigenvectors than other selection procedure considered. Mi-Lasso is also over 60 times faster than the alternative selection procedures for this application.

The rest of this paper is organised as follows, Section \ref{sec:model1} describes the underlying model. Section \ref{sec:SF} discusses the statistical aspects of ESF and looks at existing methods for the ESF eigenvector selection problem. Section \ref{sec:MiLasso} presents the Mi-Lasso procedure and derives several theoretical results. Section \ref{sec:sim1} provides a Monte Carlo study comparing Mi-Lasso to the main existing selection procedures. Section \ref{sec:app1:bh} tests the proposed method in an empirical application on house prices. Finally, Section \ref{conc1} offers our concluding remarks.

\section{Underlying model}\label{sec:model1}

Consider the following equation, where the endogenous $n\times 1$ vector $\bm{y}$ is specified as a function of an $n\times k$ matrix of exogenous regressors $\bm{X}$ and follows some spatial process:
\begin{equation}\label{model}
  \bm{y} =\bm{X\beta}_0+f(\bm{W,y,X,r})+\bm{v},
\end{equation}
where $\bm{\beta}_0$ is the $k \times 1$ parameter vector of interest and $f(\bm{W,y,X,r})$ is a linear-in-parameter function of an $n\times n$ SWM of known constants $\bm{W}$,\footnote{We allow for the $\bm{W}$ to be normalised by a scalar factor as it allows for the recovery of the original autoregressive parameters \citep{kelpr10} and maintains symmetry.} $\bm{y}$, $\bm{X}$ and an $n\times 1$ vector $\bm{r}$. One example of such a model is:
\begin{align}
    \bm{y} &=\sum^p_{i=1}\bm{W}^i\bm{y}\rho_{i,0}+\bm{X\beta}_0+ \bm{WX}\bm{\psi}_0+\bm{r}, \label{sf_sdmara}\\
    \bm{r} & = \delta_0 \bm{Wr} +\bm{v}, \label{sf_sdmarb}
\end{align}
where $\bm{\psi}_0$, $\rho_{i,0}$'s and $\delta_0$ describe the degree of spatial correlation in each of the $k$ exogenous variables, the dependent variable and error term. Note, simpler spatial models can be recovered by setting the spatial parameters $\rho_{i,0}$'s, $\delta_0$, and/or $\bm{\psi}_0$ equal to zero, and most spatial models set $p=1$. If the DGP of $\bm{y}$ is \eqref{sf_sdmara} and \eqref{sf_sdmarb} then the reduced form for $\bm{y}$ is
\begin{equation*}
    \bm{y}=\bm{S}_1^{-1}(\bm{X}\bm{\beta}_0+\bm{WX}\bm{\psi}_0+\bm{S}_2^{-1}\bm{v}),
\end{equation*}
if both $\bm{S}_1\equiv(\bm{I}-\sum^p_{i=1}\bm{W}^i\rho_{i,0})$ and $\bm{S}_2\equiv(\bm{I}-\delta_0\bm{W})$ are non-singular.

The SWM $\bm{W}$, with typical element $w_{ij}$, describes the spatial or socio-economic relationship between the cross-sectional units. When $w_{ij}\neq 0$, there is a meaningful interaction of units $j$ on unit $i$. In such cases, unit $j$ is often referred to as a neighbour of unit $i$. These interactions can stem from various sources, such as spillovers, externalities, geographic location, regulations, technology, government policy, or government expenditure. We further assume $\min_i\sum_{j=1}^nw_{ij}>0$ with probability 1, $w_{ii}=0$ by construction and $w_{ij}=w_{ji}$.
The variables $\bm{Wr}$, $\bm{WX}$ and $\bm{W}^i\bm{y}$ are typically referred to as first order spatial lags of $\bm{r}$ and $\bm{X}$ and $i$th order spatial lags of $\bm{y}$.

Let $N$ denote the set of observations $N_n=N=\{1,\ldots,n\}$. All variables are normalised, as the transformed model is estimated by a Lasso-based procedure. For reasons of generality, we allow the elements of $\bm{u}_n$, $\bm{y}_n$, $\bm{W}_n$ and $\bm{X}_n$ to be dependent on $n$, that is to form triangular arrays, however, to simplify the notation we omit the $n$ index. Our analysis is conditioned on realised values of $\bm{X}$ and  $\bm{W}$.  We consider higher-order spatial lags only as powers of the SWM $\bm{W}$ and we allow the number of lags $p$ to be unknown.\footnote{More recent papers studying the estimation of higher-order spatial models, have generalised the concept of a higher-order spatial lag to allow for $p$ different weights matrices, thus, replacing $\bm{W}^i$ with $\bm{W}_i$ in \eqref{sf_sdmara}. Powers of $\bm{W}$ are viewed as a special case. Some examples are \citet{ll10_HOgmm, be13_HOgmm, gr15_HOpinc, gr18_HOml, g19_HOstoW, bde20_HOpanel, hlx21_HObay, g18_ho, g21_HOeff, gq22_spectest}.} Even if $p$ is known, the estimation of such a model is non-trivial, as shown by \citet{b85_RCR}. When the SWM is binary, powers of the SWM can result in the presence of circular and redundant routes. Proper higher-order spatial lags need to have these circular and redundant routes eliminated.\footnote{Both \citet{bk92_algRCR} and \citet{as96_algRCR} introduced algorithms to construct proper higher-order spatial lags.}

We now make the following assumptions about variables in Equation \eqref{model}
\begin{assump}\label{as:spatial}
$\;$
\begin{enumerate}
  \item (a) $\bm{W}$ are stochastic real symmetric $n\times n$ matrices with $w_{ii}=0$. (b) The sequence $\{\bm{W}\}$
   is uniformly bounded in both row and column sums.
  \item The $n\times k$ matrices of exogenous variables $\bm{X}$ has full column rank (for large enough $n$) and all the elements of $\bm{X}$ are uniformly bound in absolute value for all $n$.
  \item The elements of the vector of innovations $\bm{v}$ are identically and independently distributed (\textit{i.i.d.}) sub-Gaussian triangular arrays with $\E[\bm{v}]=0$ and $\E[\bm{v}\bm{v}']=\sigma^2_{v}\bm{I}$ where $0<\sigma^2_{v}<\infty$. Additionally, the innovation's fourth moment is assumed finite.
\end{enumerate}
\end{assump}

Assumption \ref{as:spatial}.1-\ref{as:spatial}.3 are standard assumptions in the spatial econometrics literature \citep{kelpru98,kp99, lee04}.
Assumption \ref{as:spatial}.1 (a) is required for the spectral decomposition. Assumption \ref{as:spatial}.1 and (b) is necessary to limit the degree of dependence in $\bm{y}$. Given Assumption \ref{as:spatial}.1 (a) if the true model is \eqref{sf_sdmara}-\eqref{sf_sdmarb} and $\bm{W}$ is normalised by the largest eigenvalue then invertibility of $\bm{S}_1$ and $\bm{S}_2$ holds if
 $\sum^p_{i=1}|\rho_{i,0}|<1$ and $|\delta_0|<1$. Assumption \ref{as:spatial}.2 ensure that the Gram matrix $\bm{X}'\bm{X}/n$ is invertible.
Assumption \ref{as:spatial}.3 requires the errors to be sub-Gaussian, this assumption allows us to derive a probability for the Lasso tuning parameter dominating the noise of the model. The finite fourth moment is needed for the selection consistency proof.

\section{Eigenvector Spatial Filtering}\label{sec:SF}

\subsection{Spectral Decomposition and Spatial Filtering}\label{sec:ESF}

We now show how eigenvectors from a spectral decomposition of $\bm{W}$ can be used to spatially filter the model described in Section \ref{sec:model1}. As $\bm{W}$ is a real and symmetric matrix (by Assumption \ref{as:spatial}.1 (a)) the spectral decomposition of $\bm{W}$ is given by
\begin{equation}
  \bm{W} =\bm{E\Lambda E}', \label{specD}
\end{equation}
where $\bm{E}$ is an $n\times n$ matrix of the $n$ eigenvectors $\bm{e}_{i\in N}$ and $\bm{\Lambda}$ is a $n\times n$ diagonal matrix of the $n$ eigenvalues ($\lambda_{i\in N}$) from $\bm{W}$. It is also important to note that the matrix of eigenvectors $\bm{E}$ of $\bm{W}$ is also the matrix of the eigenvectors of $\bm{W}^i \;\; \forall \; i\in \mathbb{Z}^+$. The proof is very simple, multiplying \eqref{specD} by $\bm{W}$ and using the orthogonal nature of the eigenvectors $\bm{E}$ to substitute $\bm{E}'\bm{E} = \bm{EE}' = \bm{I}$:
\begin{equation}\label{Wpower}
  \bm{W}\bm{W}=\bm{W}^2 = \bm{W}\bm{E}\bm{\Lambda}\bm{E}' = \bm{E}\bm{\Lambda}\bm{E}'\bm{E}\bm{\Lambda}\bm{E}' = \bm{E}\bm{\Lambda}^2\bm{E}'
\end{equation}
Recursive application of \eqref{Wpower} for any $p$ results in $\bm{W}^p=\bm{E}\bm{\Lambda}^p\bm{E}'$.

The intuition behind ESF is to use individual eigenvectors $\bm{e}_{i\in N}$ as explanatory variables to proxy for $f(\bm{W,y,X,r})$, yielding a high dimensional reduced form model:
\begin{equation}\label{fullESF}
    \bm{y}\approx \bm{X\beta}_0 + \bm{E\gamma}_0 + \bm{v},
\end{equation}
where $\bm{E\gamma}_0$ can be viewed as a linear approximation of $f(\bm{W,y,X,r})$. The key problem with \eqref{fullESF} is that it is ghigh-dimensional and cannot be estimated consistently by OLS as the assumption that the regressor matrix $\bm{G}=[\bm{X},\bm{E}]$ has full column rank is violated.\footnote{This is because of $\rank(\bm{G})=\rank(\bm{G}'\bm{G})\leq \min(n,(n+k))$.} This implies a rank-deficient Gram matrix $\bm{\bm{G}'\bm{G}}/n$ with zero-valued eigenvalues. To handle this problem, we make the following assumptions:
\begin{assump}\label{as:aprx}
 $\;$
 \begin{enumerate}
   \item $||\bm{\gamma}_0||_0=s<n-k$ where $s=s_n$ is the cardinality of the active set $\Omega:=\supp(\bm{\gamma}_0)$.
   \item $f(\bm{W,y,X,r})\approx\bm{E}\bm{\gamma}_0=\bm{E}_\Omega\bm{\gamma}_\Omega$ where $\bm{E}_{\Omega}$ is an $n\times s$ matrix with columns that correspond to $\Omega$ and $\bm{\gamma}_\Omega$ the corresponding vector of unknown constants.
 \end{enumerate}
\end{assump}

Assumption \ref{as:aprx}.1 is a weak sparsity assumption, and Assumption \ref{as:aprx}.2 is required for the ESF approximation to be valid. While strong and untestable, they formalise the intuition of \citet{grif2000,grif03}, who argue only a specific subset of eigenvectors ($\bm{E}_\Omega$) are related to the dependent variable $\bm{y}$ and will have non-zero coefficients. These assumptions imply \eqref{fullESF} can be reduced to the following low-dimensional equation, where $\bm{\Upsilon}_0=[\bm{\beta}_0,\bm{\gamma}_\Omega]'$ and $\bm{G}_\Omega=[\bm{X},\bm{E}_{\Omega}]$.
\begin{align}
   \bm{y}\approx \bm{G}_\Omega \bm{\Upsilon}_0 + \bm{v}, \label{sf}
\end{align}

In principle, \eqref{sf} can be estimated by OLS. However, as $\bm{E}_{\Omega}$ is unknown, this is infeasible in practice. Thus, we now have a selection problem.

\subsection{Relationship between Moran's $I$ and ESF}\label{sec:mI}

The ESF method of \citet{grif2000} is based on the Moran's $I$ statistic for spatial autocorrelation \citep{moran50}. The test statistic for the Moran's $I$ ($m$) on the regression residual $\bm{M_Xy}=\hat{\bm{u}}$ of $\bm{y}=\bm{X\beta}+\bm{u}$ where $\bm{M_X}=\bm{I}-\bm{X}(\bm{X}'\bm{X})\bm{X}'$ is given by:
\begin{equation}\label{mIres1}
       m = \frac{\bm{y}'\bm{M_XWM_Xy}}{\bm{y}'\bm{M_Xy}}=\frac{\hat{\bm{u}}'\bm{W}\hat{\bm{u}}}{\hat{\bm{u}}'\hat{\bm{u}}},
\end{equation}
where $\bm{W}$ a $n\times n$ real symmetric SWM.\footnote{The assumption of symmetry of the elements of $\bm{W}$ is maintained \textit{w.l.o.g.} since $\hat{\bm{u}}'\bm{W}\hat{\bm{u}} = \hat{\bm{u}}'[(\bm{W}+\bm{W}')/2]\hat{\bm{u}}$ \citep{kelpru01mI}.} Substituting \eqref{specD} in \eqref{mIres1}:
\begin{equation*}
    m =  \frac{\hat{\bm{u}}' \bm{E\Lambda E}\hat{\bm{u}}}{\hat{\bm{u}}'\hat{\bm{u}}}
\end{equation*}

\citet{dejong84} showed that range of $m$ is determined by the maximum and minimum eigenvalues of $\bm{M_XWM_X}$. \citet{tiefboot95} showed that each of the $n$ eigenvalues of this expression represents a distinct $m$ values and all other possible $m$ values are just linear combinations of these $n$ values \citep{boot00}.

It is important to note that the numerator of $m$ includes $\bm{E}'\hat{\bm{u}}$, which given the orthogonality of eigenvectors is the OLS coefficient estimate from a regression of $\hat{\bm{u}}$ on $\bm{E}$.\footnote{In other words, $\bm{E}'\hat{\bm{u}} = (\bm{E}'\bm{E})^{-1}\bm{E}'\hat{\bm{u}}$} \citet{grif03} argues that each of the $n$ eigenvectors represents mutually orthogonal spatial patterns and only a subset of eigenvectors will be relevant to the model, i.e., in a regression framework only a subset of eigenvectors will have non-zero coefficients.

\subsection{Existing Selection Procedures}

Running the ESF method requires identifying $\bm{E}_\Omega$, the relevant set of eigenvectors. The first type of procedures proposed were forward stepwise greedy algorithms where eigenvectors are iteratively added until some user-specified threshold is reached \citep{grif2000,grif03,tiefgrif07, md19}. \citet{grif03} proposed iteratively adding eigenvectors in a greedy manner to the base regression
\begin{equation}
  \bm{y}=\bm{X\beta}+\bm{u}, \label{missOLSa}
\end{equation}
until the spatial correlation in the OLS residual $\hat{\bm{u}}$ falls below a pre-specified level. Selection criteria based on alternative statistics such as the adjusted-$R^2$, the Akaike Information Criterion or Bayesian Information Criterion have also been suggested \citep{tiefgrif07, md19}. \citet{tiefgrif07} specifically suggest using the standardised Moran's $I$ as the criterion for the greedy algorithm, based on its power against a wide array of autoregressive models and residual distributions \citep{ar91} and the fact it can be used for small samples \citep{kel17}. The standardised version of Moran's $I$ statistic ($Z$) on the residual $\hat{\bm{u}}$ is:\footnote{Note the matrix $\bm{X}$ in the orthogonal projection matrix $\bm{M_X}$ may also include the selected eigenvectors in \cite{tiefgrif07} procedure.}
\begin{equation}\label{zmi2}
    Z	= \Bigg( \frac{m-\E[m]}{\sqrt{\var(m)}} \Bigg)
\end{equation}
with
\begin{equation*}
    \E[m] = \frac{tr(\bm{M_XWM_{X}})}{n-k}
\end{equation*}
and
\begin{equation*}
    \var(m) = \frac{2\bigg((n-k)tr\big((\bm{M_{X}WM_{X}})^2\big)- \big[tr(\bm{M_{X}WM_{X}})\big]^2\bigg)}{(n-k)^2(n-k-2)}.
\end{equation*}

The greedy algorithm iterates over the candidate set of eigenvectors $\bm{E}_{c}$, searching for the eigenvector that minimizes $Z$. The selected eigenvector $\bm{e}_{i\in N}$ is then removed from $\bm{E}_{c}$ and added to the design matrix of \eqref{missOLSa}, and the residuals $\hat{\bm{u}}$ of this updated regression are tested to check if $|Z|< \epsilon$, where $\epsilon$ is a pre-specified threshold level of $Z$, which they suggest should be dependent on the sample size $n$.\footnote{\cite{tiefgrif07} suggest if $n<50$ then $\epsilon\approx1.0$ and if $n\approx 500$ then $\epsilon\approx0.1$.} If the condition is satisfied the iterations stop, if not the algorithm continues searching in the remaining candidate eigenvector set $\bm{E}_{c}$, with this iterative process continuing until $|Z|< \epsilon$.


\citet{grif03} argues that the candidate eigenvectors $\bm{E}_{c}$ form a subset  $\bm{E}_{c} \subseteq \bm{E}$ of the full set of eigenvectors, based on several criteria. First, if $\bm{y}$ exhibits positive global spatial autocorrelation then $\bm{E}_{c}$ should be restricted to those eigenvectors with associated positive eigenvalues, as these are associated with at least weak positive spatial autocorrelation. Second, eigenvectors with small eigenvalues should be excluded from $\bm{E}_{c}$, suggesting a minimum threshold eigenvalue of 0.25, which is related to only approximately 5\% of the variation attributed to spatial correlation in the dependent variable.

These forward stepwise procedures, through intuitive, have several key disadvantages. First, a lot of parameters are left to the user's discretion, such as, which statistic or information criterion to use, what threshold $\epsilon$ to use, which eigenvectors to include in the initial $\bm{E}_{c}$, and in which order to add the eigenvectors. Second, these greedy algorithms could also be at risk of data mining, with estimated models falling victim to over-fitting. Third, all these approaches are heuristics that aim to simplify the original, and infeasible, subset sum problem; therefore the solutions they obtain will be sub-optimal, with no guarantee they are close to the optimal one. Finally, these sequential methods carry a large computational burden, which becomes more acute when $n$ is large. This can be mitigated by limiting $\bm{E}_{c}$ with the rules of thumb mentioned above, but again with no guarantee these rules will consistently recover $\bm{E}_\Omega$.

This motivates \citet{seya15} to propose using Lasso \citep{t96}, which shrinks many of the coefficients to zero, and can thus be used for variable selection \citep{tibs09}. \citet{seya15} use Lasso under the assumption the parameter vector $\bm{\gamma}_0$ is sparse and the matrix of regressors $\bm{X}$ has full column rank, so that only the $\bm{\gamma}$ vector is penalised. The resulting Lasso estimator is:
\begin{equation}\label{Lasso1}
 [\hat{\bm{\beta}}_\theta,\hat{\bm{\gamma}}_\theta]\in \min_{ \bm{\beta}\in\R^k} \min_{ \bm{\gamma}\in\R^n} \{ ||\bm{y}-\bm{X\beta}-\bm{E\gamma}||^2_2 +\theta ||\bm{\gamma}||_1 \},
\end{equation}
where $\theta>0$ is the Lasso regularization or tuning parameter. Equation \eqref{Lasso1} defines a family of estimators indexed by the tuning parameter $\theta$, a hyperparameter that ultimately determines which eigenvectors the Lasso selects.

\citet{seya15} proposed using $k$-fold cross-validation (CV) combined with the Brent algorithm \citep{brent73} to estimate $\hat{\theta}$, with prediction accuracy as the loss function. The Brent algorithm is a root-finding algorithm that allows for the optimisation to be non-convex: the algorithm first tries inverse quadratic interpolation in an attempt to achieve faster convergence which works well if the optimisation is convex. If it is non-convex and inverse quadratic interpolation fails, (slower) linear interpolation is used instead. CV using the Brent algorithm is the most time-consuming part of the \citet{seya15} Lasso procedure. Because the theoretical results on CV-Lasso hinge on the assumption that the cross-sectional units are independent \citep{czc20}, it is hard to justify their validity for ESF, where eigenvectors are derived from a matrix that encodes cross-sectional dependence.\footnote{CV procedures do exist for cross-sectionally dependent data but they need to be carefully designed, for example see \citet{llz20}.}

Some other methods have also been proposed. \cite{lpz13sea} suggest simply including the first $j$ eigenvectors (sorted by eigenvalue magnitude) where $j$ is simply based on the sample size. Given this fixed rule, \cite{lpz13sea} finds the quality of the ESF approximation is sensitive to the underlying spatial processes. \citet{ch16ESFsel} argue more eigenvectors are needed when the level of spatial correlation is high compared to when the level of spatial correlation is low, thus, simple rules based on for example sample size may result in a sub-optimal set of eigenvectors being selected. \citet{ch16ESFsel} instead develop the following eigenvector selection rule via simulation:
\begin{equation}\label{chunsel}
    w= \frac{n_{pos}}{1+\exp\big[2.1480-\big(6.1808(m+0.6)^{0.1742}\big)\big/n_{pos}^{0.1298}+3.3534\big/(m+0.6)^{0.1742}\big]},
\end{equation}
where $n_{pos}$ denotes the number of eigenvectors that exhibit positive spatial correlation (eigenvectors with positive eigenvalues). Equation \eqref{chunsel} was generated from a limited simulation that assumed the DGP has just spatial autoregressive disturbances, \citet{ch16ESFsel} do not evaluate how their rule performs when the DGP follows some other spatial process.

\section{Theoretical properties of Moran's $I$ Lasso}\label{sec:MiLasso}
\subsection{Moran's $I$ Lasso framework for eigenvector selection}

The Lasso estimates are ultimately determined by tuning parameter $\theta$. Supposing $\theta=0$, the Lasso solution reduces to the OLS solution, whereas with a sufficiently large $\theta$ the penalised parameter vector is shrunk to zero (no eigenvectors selected). More moderate values of $\theta$ will result in some parameters being shrunk towards zero and some to precisely zero. As outlined above, the goal of ESF is to eliminate spatial correlation patterns in a linear regression framework. Information about these patterns will be contained in the regression residuals $\hat{\bm{u}}$, and we propose using these to determine a point estimate for $\theta$.

\begin{algorithm}[t]
\caption{Mi-Lasso Algorithm}
\label{MiLassoalg}
\begin{enumerate}
  \item Decompose the SWM to get the candidate set of Eigenvectors $\bm{E}$.
  \item Estimate simple residuals $\hat{\bm{u}}= \bm{M}_{X} \bm{y}$ where $\bm{M}_X = \bm{I}-\bm{X}(\bm{X}'\bm{X})^{-1}\bm{X}'$
  and calculate corresponding the absolute standardised Moran's $I$ of $\hat{\bm{u}}$ denoted $Z$
    \item Estimate
    \begin{equation}\label{Lasso1a}
   [\hat{\bm{\beta}},\hat{\bm{\gamma}}]\in \min_{ \bm{\beta}\in\R^k} \min_{ \bm{\gamma}\in\R^n} \{ ||\bm{y}-\bm{X\beta}-\bm{E\gamma}||^2_2 +|\frac{1}{Z^2}|\cdot ||\bm{\gamma}||_1 \}
  \end{equation}
   Use the Lasso or post-Lasso estimates of \eqref{Lasso1a}
\end{enumerate}
\end{algorithm}

It seems reasonable to assume that when the level of spatial correlation in the residuals is low, only a small set of eigenvectors is necessary. Thus, a high level of regularization (value of $\theta$) is required. In contrast, when the level of spatial correlation is high, a large set of eigenvectors will be necessary. Thus, a low level of regularization (value of $\theta$) is required. Following \citet{tiefgrif07} we propose using the standardised Moran's $I$ \eqref{zmi2} to measure the spatial correlation of the residuals due to the previously mentioned properties. As $Z$ takes on large values when the correlation is high and small values when the correlation is low, we propose using the inverse of the square of $Z$ from the residuals of \eqref{missOLSa} as a point estimate of $\theta$,
\begin{equation}\label{mytheta}
  \theta = \frac{1}{Z^2}, \qquad \forall \enspace Z \neq 0
\end{equation}
The square is chosen to ensure the tuning parameter is always positive.\footnote{A positive tuning parameter is necessary to ensure Lasso gives a unique solution.} The proposed estimator is called Moran $I$ Lasso (Mi-Lasso) and is outlined in Algorithm \ref{MiLassoalg}.

As Lasso is a shrinkage estimator, it induces a downward bias on the estimated non-zero coefficients. Post-Lasso (pLasso) uses the Lasso estimator as selection procedure (assuming Lasso selects the correct variables), and then OLS is applied to the model selected by Lasso, straightforwardly providing unbiased estimates and standard errors.\footnote{For formal results on Post-Lasso, see \citet{bell13}.} The Morans' $I$ Post-Lasso (Mi-pLasso) estimator is defined as:
\begin{equation*}\label{pLasso}
 [\hat{\bm{\beta}},\hat{\bm{\gamma}}]= \min_{ \bm{\beta}\in\R^k} \min_{ \bm{\gamma}\in\R^n}  ||\bm{y}-\bm{X\beta}-\bm{E\gamma}||^2_2 \;\;\; \st \;\;\; \supp(\bm{\gamma}_0)=\supp(\hat{\bm{\gamma}}_\frac{1}{Z}).
\end{equation*}


To focus the theoretical analysis on the parameter vector $\bm{\gamma}$, we use the Frisch-Waugh-Lowell (FWL) partial regression theorem to partial out the $\bm{X}$ matrix. \citet{tibt11} and \citet{Lassofwl} show that the FWL theorem could be used in a low-dimensional Lasso setting. Lemma \ref{lem:hdfwl} shows that the FWL theorem can also be applied to the high-dimensional case of Mi-Lasso.

\begin{lem}\label{lem:hdfwl}
 Consider the following two Lasso regressions:
 \begin{align}
   [\hat{\bm{\beta}},\hat{\bm{\gamma}}] &= \min_{ \bm{\beta}\in\R^k} \min_{ \bm{\gamma}\in\R^n} \{ ||\bm{y}-\bm{X\beta}-\bm{E\gamma}||^2_2 +\frac{1}{Z^2} ||\bm{\gamma}||_1 \},  \label{Lassofull} \\
   [\tilde{\bm{\gamma}}]&= \min_{ \bm{\gamma}\in\R^n} \{ ||\tilde{\bm{y}}-\tilde{\bm{E}}\gamma||^2_2 +\frac{1}{Z^2} ||\bm{\gamma}||_1 \},  \label{Lassofwl}
\end{align}
where $\bm{X}$ is an $n\times k$ matrix, $\bm{E}$ is an $n\times n$ matrix, $\tilde{\bm{y}}=\bm{M_Xy}$, $\tilde{\bm{E}}=\bm{M_XE}$ with $\bm{M_X}= \bm{I}- \bm{X}(\bm{X}'\bm{X})^{-1}\bm{X}'$.
Then if Assumption \ref{as:spatial}.2 holds $\hat{\bm{\gamma}}=\tilde{\bm{\gamma}}$
\end{lem}

The proof is provided in appendix \ref{app:proofs}.

We now introduce the following additional notation in the design. Without loss of generality, let $\bm{C}_{\Omega\Omega}=n^{-1}\tilde{\bm{E}}_{\Omega}'\tilde{\bm{E}}_{\Omega}$, $\bm{C}_{\Omega\grave{\Omega}}=n^{-1}\tilde{\bm{E}}_{\Omega}'\tilde{\bm{E}}_{\grave{\Omega}}$, $\bm{C}_{\grave{\Omega}\Omega}=n^{-1}\tilde{\bm{E}}_{\grave{\Omega}}'\tilde{\bm{E}}_{\Omega}$ and $\bm{C}_{\grave{\Omega}\grave{\Omega}}=n^{-1}\tilde{\bm{E}}_{\grave{\Omega}}'\tilde{\bm{E}}_{\grave{\Omega}}$
where $\tilde{\bm{E}}_{\Omega}$ is an $n\times s$ matrix with columns corresponding to the active set $\Omega$. $\grave{\Omega}$ is the complement set and the $n\times q$ matrix $\tilde{\bm{E}}_{\grave{\Omega}}$ is defined accordingly with $q_n=q=s-n$. Now the (re-scaled) Gram matrix $\bm{C}_n=\bm{C}=n^{-1}\tilde{\bm{E}}'\tilde{\bm{E}}$ can be expressed in block-wise form as:
\begin{equation*}
 \bm{C}  = \begin{bmatrix} \bm{C}_{\Omega\Omega} & \bm{C}_{\Omega\grave{\Omega}} \\   \bm{C}_{\grave{\Omega}\Omega} & \bm{C}_{\grave{\Omega}\grave{\Omega}} \end{bmatrix}.
\end{equation*}
Similarly we define $\bm{\gamma}=[\bm{\gamma}_{\Omega},\bm{\gamma}_{\grave{\Omega}}]'=[\gamma_1,\ldots,\gamma_{s},\gamma_{s+1},\ldots,\gamma_{n}]'$.

\subsection{Non-asymptotic bounds}

This section produces performance bounds for the Mi-Lasso estimates of $\bm{\gamma}$. Given the high-dimensional structure of ESF, the Gram matrix $\bm{G}'\bm{G}/n$ is singular. This implies its minimum eigenvalue will be zero. However, as shown by \citet{lasdig09} for the case of Lasso, the following restricted eigenvalue (RE) condition only requires the appropriate sub-matrix of the Gram matrix to have positive and finite eigenvalues.
\begin{assump}\label{as:re}
 Let $\bar{b}$ and $t$ be positive constants and $\Omega$ denote the active set. Then the restricted eigenvalue condition holds for $\tilde{\bm{E}}$, as $n\to \infty$ if we assume:
 \begin{equation}\label{RE}
   \tau_{min} := \min_{\mathcal{C}(\Omega,\bar{b})} \frac{||\tilde{\bm{E}}\bm{\Delta}||_2}{\sqrt{n}||\bm{\Delta}||_2}\geq t >0,
 \end{equation}
 where
 \begin{equation}\label{rset}
   \mathcal{C}(\Omega,\bar{b}) =\{\bm{\Delta}\in \R^n:||\bm{\Delta}_{\grave{\Omega}}||_1 \leq \bar{b}  ||\bm{\Delta}_{\Omega}||_1, \enspace \delta \neq 0\}
 \end{equation}
 and $\bm{\Delta}=\tilde{\bm{\gamma}}-\bm{\gamma}_0$.
\end{assump}

Assumption \ref{as:re} requires that $\bm{\Delta}$ lies within the restricted set \eqref{rset}. As $\bm{\Delta}$ is the difference between the estimate $\tilde{\bm{\gamma}}$ and the true parameter $\bm{\gamma}_0$, the restricted eigenvalue bounds the minimum change in the prediction norm from a deviation $\bm{\Delta}$ within the restricted set $\mathcal{C}(\Omega,\bar{b})$ relative to the norm of the deviation on the true support $\bm{\Delta}_{\Omega}$.

By combining Assumptions \ref{as:spatial} and \ref{as:aprx} with the RE condition, and treating $\bm{X}$ and $\bm{E}$ as constants (realisations) we can now establish the $\ell_1$ and $\ell_2$ parameter norm bounds and the $\ell_2$ prediction norm bound for the Mi-Lasso estimates of $\bm{\gamma}$.
\begin{theorem}\label{the:parcon}
Suppose Assumption \ref{as:spatial}-\ref{as:aprx} and Assumption \ref{as:re} holds for $\bar{b}=\frac{b+1}{b-1}$ for some $b\geq 1$ and the regularization parameter satisfies $\frac{1}{Z^2} \geq b2\sqrt{\frac{4\sigma^2_{\bm{v}}\log n}{n}}$ with probability tending to one as $n\to \infty$, then:
\begin{equation}\label{parcon1}
  ||\tilde{\bm{\gamma}}-\bm{\gamma}_0||_1 \leq \frac{\big(\frac{1}{b}+1\big)s }{\tau_{min}^2Z^2n},
\end{equation}
\begin{equation}\label{parcon2}
  ||\tilde{\bm{\gamma}}-\bm{\gamma}_0||_2 \leq \frac{\big(\frac{1}{b}+1\big)\sqrt{s} }{\tau_{min}^2Z^2n},
\end{equation}
\begin{equation}\label{parcon3}
  \frac{1}{\sqrt{n}}||\tilde{\bm{E}}(\tilde{\bm{\gamma}}-\bm{\gamma}_0)||_2 \leq \frac{\big(\frac{1}{b}+1\big)\sqrt{s} }{\tau_{min}Z^2n}.
\end{equation}
\end{theorem}

The proof is provided in appendix \ref{app:proofs}.

The three convergence rates presented in Theorem \ref{the:parcon} depend on the number of eigenvectors with non-zero coefficients, the sample size, and $Z$. They also require that the tuning parameter dominates the noise of the model. By assuming the errors are sub-Gaussian (Assumption \ref{as:spatial}.3) we prove the probability of this event occurring goes to one as $n\to \infty$ (see proof for further details).

\begin{corollary}\label{coro1}
If the condition of Theorem \ref{the:parcon} are satisfied and $s/Z^2n=o_p(1)$ then the bounds \eqref{parcon1}-\eqref{parcon3} are $o_p(1)$ as $n\to \infty$.
\end{corollary}

Corollary \ref{coro1} is satisfied if $Z=O_p(1)$,, which is reasonable as $Z$ is a measure of correlation, and $s$ grows at a rate slower than $n$, which is satisfied by Assumption \ref{as:hidem}.4 below.

\subsection{Consistent Eigenvector Selection}

This section shows the conditions required for Mi-Lasso to consistently select the non-zero and zero elements in $\bm{\gamma}$. Following \citet{zhao06}, we say that $\tilde{\bm{\gamma}} =_s \bm{\gamma}_0$ if and only if $\sign(\tilde{\bm{\gamma}}) = \sign(\bm{\gamma}_0)$ where $\sign(\cdot)$ maps positive entry to 1, negative entry to -1 and zero to zero. We now define selection consistency for Mi-Lasso as
\begin{defin}\label{def:con}
\citep{zhao06} Mi-Lasso estimates of $\bm{\gamma}$ are selection consistent if:
\begin{equation*}
 \lim_{n\to\infty}P(\tilde{\bm{\gamma}}=_s \bm{\gamma}_0) =1.
\end{equation*}
\end{defin}

The following assumptions are required to prove sign consistency of Mi-Lasso.
\begin{assump}\label{as:hidem}
    There exists $M_1,M_2,M_3 > 0$, $0\leq c_1<c_2 \leq 1$ and a vector of postive constants $\bm{\nu}$, the following holds:
  \begin{enumerate}
    \item \qquad $\frac{1}{n}\tilde{\bm{e}}_{i}'\tilde{\bm{e}}_{i} \leq M_1 \;\; \forall i,$

    \item \qquad $\bm{\alpha}'\bm{C}_{\Omega\Omega}\bm{\alpha} \geq M_2 \;\; \forall \; ||\bm{\alpha}||_2^2=1,$
    \item \qquad $n^{\frac{1-c_2}{2}}\min_{i=1,\ldots,s}|\gamma_{i}|\geq M_3,$
    \item \qquad $s = O(n^{c_1}),$
    \item \qquad $ |\bm{C}_{\grave{\Omega}\Omega}(\bm{C}_{\Omega\Omega})^{-1}\sign(\bm{\gamma}_{\Omega})|\leq \bm{1}-\bm{\nu}.$
  \end{enumerate}
\end{assump}


Assumption \ref{as:hidem}.1 is a normalisation of the transformed eigenvectors. Assumption \ref{as:hidem}.2 bounds the eigenvalue of the eigenvectors with non-zero coefficients from below, so the inverse of $\bm{C}_{\Omega\Omega}$ is well behaved. Assumption \ref{as:hidem}.3 and Assumption \ref{as:hidem}.4 are important as they ensure convergence in the high dimensional space as $n\to\infty$. Assumption \ref{as:hidem}.3 ensure there is a difference of size $n^{c_2}$ between the decay rate of $\bm{\gamma}_{\Omega}$ and $\sqrt{n}$, preventing the estimates from being dominated by the disturbance terms, which aggregate at a rate of $n^{-1/2}$. Assumption \ref{as:hidem}.4 is a sparsity assumption that requires the square root of the size of the true model $\sqrt{s}$ to increase at a slower rate than the rate difference, preventing the Lasso estimation bias from dominating the model parameters. Assumption \ref{as:hidem}.5 (assuming $\bm{C}_{\Omega\Omega}$ is invertible) is the Irrepresentable Condition (IC), which is the necessary condition for the consistency of Mi-Lasso selection, the inequality holds element-wise. The IC requires the correlation between the relevant and irrelevant eigenvectors to be zero or weak. In the Mi-Lasso framework, this is likely to be satisfied as the columns of $\bm{E}$ are mutually orthogonal. The columns of $\tilde{\bm{E}}$ may not be, however, as the eigenvectors are projected into the column space of $\bm{X}$. Unfortunately, in practice, the IC is impossible to verify as we do not know the true parameter vector $\bm{\gamma}_0$.

The following proposition places a lower bound on the probability of Mi-Lasso picking the true model, which quantitatively relates to the probability of Lasso selecting the correct model. Proposition \ref{prop1} is a modification of Proposition 1 in \citet{zhao06}.
\begin{prop}\label{prop1}
  Assume Assumption \ref{as:spatial}, \ref{as:aprx} and \ref{as:hidem}.5 holds for some $\bm{\nu}>0$, then:
  $$\p\big(\tilde{\bm{\gamma}}=_s\bm{\gamma}_0\big)\geq \p(A \cap B),$$
  for
  \begin{align*}
    A = & \{||(\bm{C}_{\Omega\Omega})^{-1}\bm{z}_{\Omega}||<\sqrt{n}(|\bm{\gamma}_{\Omega}|-\frac{1}{2Z^2n}||(\bm{C}_{\Omega\Omega})^{-1}\sign(\bm{\gamma}_{\Omega})||)\}, \\
    B = & \{||\bm{C}_{\grave{\Omega}\Omega}(\bm{C}_{\Omega\Omega})^{-1}\bm{z}_{\Omega}-\bm{z}_{\grave{\Omega}}||\leq \frac{1}{2Z^2\sqrt{n}}\bm{\nu}\},
  \end{align*}
  where $\bm{z}_{\Omega}=\frac{1}{\sqrt{n}}\tilde{\bm{E}}_{\Omega}'\bm{v}$ and $\bm{z}_{\grave{\Omega}}=\frac{1}{\sqrt{n}}\tilde{\bm{E}}_{\grave{\Omega}}'\bm{v}$.
\end{prop}

The proof is provided in appendix \ref{app:proofs}.

Proposition \ref{prop1} shows that the measure of spatial correlation $Z$ determines the size of the trade-off between events $A$ and $B$. A higher level of spatial correlation will lead to larger $A$ but smaller $B$; this makes Mi-Lasso more likely to select irrelevant eigenvectors. In contrast, a larger $\nu_i$ has no impact on $A$ but leads to a larger $B$. So when IC holds with a large $\nu_i$, Mi-Lasso is more likely to select the correct model.
\begin{theorem}\label{the:selcon}
  Assuming Assumption \ref{as:spatial}, \ref{as:aprx} and \ref{as:hidem} hold, and $c_2-c_1=0.5$. Given $s+q=n$ implies Mi-Lasso is sign consistent for all $\frac{1}{Z^2}$ that satisfy $\frac{1}{Z^2\sqrt{n}}= o_p(n^{\frac{c_2-c_1}{2}})= o_p(n^{\frac{1}{4}})$ and $\frac{1}{n^3Z^{8}}\to \infty$,
  we have
  $$\p\big(\tilde{\bm{\gamma}}=_s\bm{\gamma}_0\big)\geq 1 -O(n^3Z^{8}) \to 1 \;\;\; as \; n\to \infty.$$
\end{theorem}

The proof is provided in appendix \ref{app:proofs}.

Theorem \ref{the:selcon} shows that Mi-Lasso is consistent in selecting the true model if the 4$th$ moment of the errors is finite (Assumptions \ref{as:spatial}.3), Assumptions \ref{as:spatial}-\ref{as:hidem} hold and the difference between $c_2$ and $c_1$ is 0.5. The greatest difference (between $c_2$ and $c_1$) for which Mi-Lasso is consistent is 0.5, smaller differences can also yield consistency, but this would require higher order moments of the errors to be finite. For example, if we assume the 6th or 8th moment is finite, the difference would need to be $1/3$ or 0.25 for Mi-Lasso to be consistent (see proof for further details).

\section{Monte Carlo Study}\label{sec:sim1}


To evaluate the finite sample performance of Mi-Lasso and compare it to the main existing selection procedures, we conduct two Monte Carlo exercises where the DGP is,
\begin{align}\label{dgpsim}
 \bm{y}=&\sum^p_{i=1}\bm{W}^i\bm{y}\rho_i + \beta\bm{x} + \psi \bm{Wx}+\bm{v},  \\
 \bm{x}\sim &N(0,\bm{I}), \; \bm{v}\sim N(0,\bm{I}). \nonumber
\end{align}

%

In both simulations, we set the `true' parameter value of $\beta = 1$ and $\psi=0.9$. The elements of $\bm{W}$, denoted $w_{ij}$, are independent draws from a Bernoulli distribution with success probability $\mu/n$ for some constant $\mu < \infty$, $w_{ii}=0$ and $w_{ij}=w_{ji}$. By construction, $\mu$ is the expected number of links for each unit, and we set $\mu\in\{4,8,12\}$.
Each $\bm{W}$ is normalised by the maximal of the row (or column) sum. Sample sizes considered are $n\in\{100,250,500\}$, and we run 1000 replications.
\begin{figure}[t]
  \caption{Bias and MSE of $\beta$ and number of selected eigenvectors, setup A with $\mu=12$}
  \label{fig:mu12}
  \centering
    \includegraphics[width=0.9\textwidth]{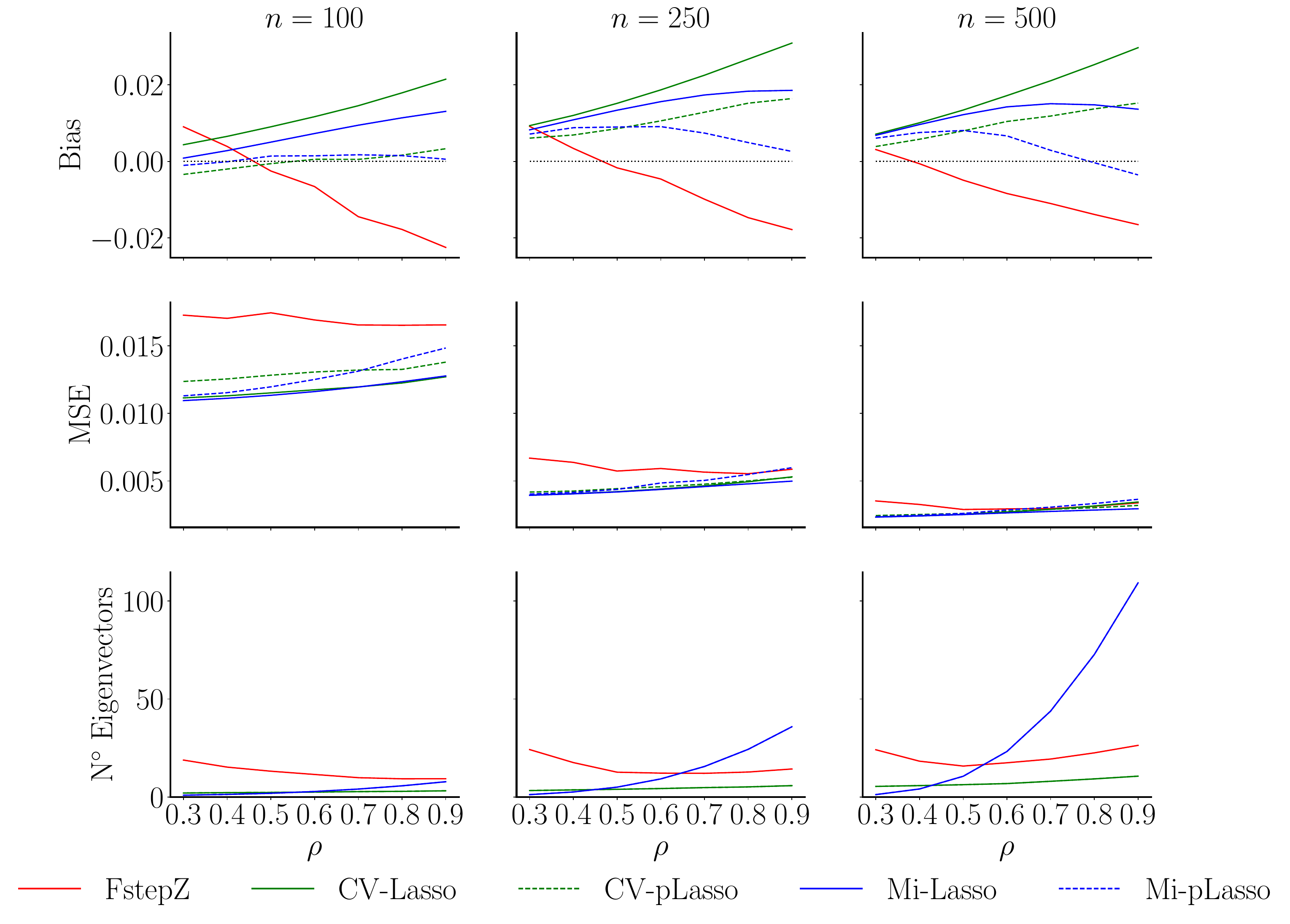}
\end{figure}

In setup A, we set $p=1$ so we can evaluate how the method performs with different levels of spatial correlation $\rho_1\in \{0.3,0.4, 0.5, 0.6, 0.7, 0.8,0.9\}$. We consider only positive spatial correlation as this is the most common setting. In setup B, we set $p=3$ to evaluate the performance of ESF in the presence of higher-order spatial lags. In both setups the estimators compared are:\footnote{An oracle estimator is not possible hare as this requires knowledge of $\supp(\bm{\gamma}_0)$, which is unknown.}
\begin{itemize}
  \item Mi-Lasso - Algorithm \ref{MiLassoalg} with step 3 using Lasso.
  \item Mi-pLasso - Algorithm \ref{MiLassoalg} with step 3 using post Lasso (OLS with the selected eigenvector)
  \item CV-Lasso - Lasso algorithm outlined in \citep{seya15}
  \item CV-pLasso - OLS with the selected eigenvector from CV-Lasso
  \item FstepZ - forward stepwise algorithm outlined in \citep{tiefgrif07} with a stopping rule $z=0.1$.
\end{itemize}

Figures \ref{fig:mu12}, \ref{fig:mu4} and \ref{fig:mu8} show the bias, MSE, and the number of selected eigenvectors for setup A,\footnote{Figures \ref{fig:mu4} and \ref{fig:mu8} are provided in appendix \ref{app:figures}} revealing the different selection behaviours of these estimators. For CV-Lasso the number of selected increases very slightly as the levels of spatial correlation in the dependent variable increases, and this pattern is consistent across different sample sizes and $\mu$. In contrast, FstepZ selects more eigenvectors when the spatial correlation level is low than high for small sample sizes and the largest set of eigenvector when the level of spatial correlation in the dependent variable is small. Mi-Lasso behaviour is as expected from the intuition of the procedure, selecting a small set of eigenvectors when the level of spatial correlation is low and a large set when the level is high.

\begin{table}[t]
  \caption{Bias, MSE and the number of selected eigenvectors for setup B}
  \label{tab:ho}
\centering
\begin{tabular}{clccccccc}
  \hline
\T  \B   &  & \multicolumn{2}{c}{$\mu= 4$} & \multicolumn{2}{c}{$\mu=8$} & \multicolumn{2}{c}{$\mu=12$}  \\
\cline{3-8}
\T \B n & Estimator & $\beta$ & $|\tilde{\bm{E}}_{\Omega}|$ & $\beta$ & $|\tilde{\bm{E}}_{\Omega}|$ & $\beta$ & $|\tilde{\bm{E}}_{\Omega}|$ \\
  \hline
\T 100 & FstepZ & -0.025(0.018) & 9 & -0.002(0.017) & 11 & -0.01(0.015) & 10 \\
  100 & CV-Lasso & 0.055(0.017) & 5 & 0.022(0.012) & 3 & 0.021(0.012) & 3 \\
  100 & CV-pLasso & 0.019(0.017) & 5 & 0.009(0.013) & 3 & 0.007(0.013) & 3 \\
  100 & Mi-Lasso & 0.034(0.016) & 20 & 0.017(0.012) & 4 & 0.016(0.012) & 4 \\
\B  100 & Mi-pLasso & 0.014(0.018) & 20 & 0.011(0.013) & 4 & 0.008(0.013) & 4 \\
\T  250 & FstepZ & -0.007(0.006) & 14 & -0.007(0.006) & 13 & -0.003(0.005) & 12 \\
  250 & CV-Lasso & 0.046(0.007) & 7 & 0.04(0.006) & 6 & 0.029(0.005) & 4 \\
  250 & CV-pLasso & 0.026(0.006) & 7 & 0.026(0.006) & 6 & 0.02(0.005) & 4 \\
  250 & Mi-Lasso & 0.034(0.006) & 38 & 0.03(0.005) & 28 & 0.024(0.005) & 14 \\
\B  250 & Mi-pLasso & 0.018(0.007) & 38 & 0.017(0.006) & 28 & 0.015(0.005) & 14 \\
\T  500 & FstepZ & -0.013(0.003) & 25 & -0.005(0.003) & 19 & -0.004(0.003) & 17 \\
  500 & CV-Lasso & 0.044(0.005) & 13 & 0.033(0.004) & 9 & 0.027(0.003) & 7 \\
  500 & CV-pLasso & 0.024(0.004) & 13 & 0.022(0.003) & 9 & 0.019(0.003) & 7 \\
  500 & Mi-Lasso & 0.025(0.004) & 132 & 0.025(0.003) & 58 & 0.022(0.003) & 36 \\
\B  500 & Mi-pLasso & 0.009(0.004) & 132 & 0.011(0.004) & 58 & 0.011(0.003) & 36 \\
   \hline
   \multicolumn{7}{l}{%
   \begin{minipage}{6cm}%
     \vspace{0.1cm}
   \footnotesize{\textit{Note}: Bias (MSE)}
     \end{minipage}%
   }\\
\end{tabular}
\end{table}

The Lasso estimators generally have a smaller bias and larger MSE than their post-Lasso (pLasso) counterparts. When the level of spatial correlation is high Mi-Lasso has the best performance in terms of bias and performs comparably to the other estimator in terms of MSE. Mi-pLasso has the smallest MSE when then level of spatial correlation is high and comparably well when the level of spatial correlation is low. Notably, FstepZ has the largest MSE when the sample size is 100 all levels of spatial correlation and $\mu$ considered and when the level of spatial correlation is low for other sample sizes. FstepZ performance in terms of bias and MSE improves as the sample size increases and the SWM becomes more dense. Generally, in terms of bias, the estimators diverge as the level of spatial correlation increases, this is because the bias is determined by an interaction between the level of $\rho_1$ and the structure of the SWM. Thus, for a given SWM, the larger $\rho_1$ the larger the bias, so mistakes/variation in selection can have a larger effect.

\begin{table}[t]
\centering
\caption{Computation Time and Sample Sizes}
\label{t:comptimeN}
\begin{tabular}{cccc}
  \hline
\B\T Sample Size & Mi-Lasso  & CV-Lasso & FstepZ \\
  \hline
\T 250 & 1 (0.09)  & 6.44 (0.58) & 101 (9.09) \\
   500 & 1 (0.15)  & 35.87 (5.38) & 1233.40 (185.01) \\
  1000 & 1 (2.13)  & 37.52 (79.92) & 321.50 (4962.64) \\
  2000 & 1 (23.49)  & 13.22 (310.71)  & 328.16 (7708.56)  \\
\B  10000 & 1 (1819.48)   & 19.38 (35258.18)  & - \\
  \hline
    \multicolumn{4}{l}{%
  \begin{minipage}{12cm}%
     \vspace{0.1cm}
\footnotesize{\textit{Note}: Relative computational time, figures in parenthesis are time in seconds. All procedures exclude eigen-decomposition and include the full set of eigenvectors in the search set. The DGP of $y$ is \eqref{dgpsim} with $p=1$, $\beta=1$, $\rho_1=0.3$ and $\mu=8$. FstepZ is the forward stepwise algorithm outlined in \citep{tiefgrif07}.}
  \end{minipage}%
}\\
\end{tabular}
\end{table}

For setup B, we set $\rho_1=0.6$, $\rho_2=0.4$ $\rho_3=0.5$. Table \ref{tab:ho} shows the bias, MSE, and the number of selected eigenvectors for setup B. This table confirms that ESF can work well in the presence of higher-order spatial lags. This table shows that Mi-Lasso selects less eigenvectors as the density of the SWM ($\mu$) increases.  Mi-Lasso and Mi-pLasso always has a smaller bias and a comparable MSE than CV-Lasso and CV-pLasso. FstepZ generally performs better in terms of both bias and MSE as the sample size increases.

Finally, Table \ref{t:comptimeN} shows the computational times of the different estimators used in the simulations. These results show Mi-Lasso is the fastest procedure, CV-Lasso is the second fastest, and FstespZ is the slowest procedure for a given sample size. Comparing Mi-Lasso to CV-Lasso, we find Mi-Lasso is up to 37 times faster. The most substantial computational gains are found when the sample size is 1000, but even when the sample size is very large (10,000), Mi-Lasso reamins 19 times faster than CV-Lasso, with FstepZ becoming unfeasible.

%

\begin{table}[ht]
    \centering
  \caption{Variables used in Boston housing application}
    \label{discr}
    \begin{tabular}{ll}
    \hline
\T \B   Variable  & Description \\ \hline
\T        p & Median values of owner-occupied housing in thousands of U.S. dollars \\
        crim & Per capita crime \\
        zn & Proportion of residential land zoned for lots over 25,000 ft$^2$ per town \\
        indus & Proportion of non-retail business acres per town \\
        cr & An indicator: 1 if tract borders Charles River; 0 otherwise \\
        nox & Nitric oxide concentration (parts per 10 million) per town \\
        rm & Average number of rooms per dwelling \\
        age & Proportion of owner-occupied units built prior to 1940 \\
        dis & Weighted distance to five Boston employment centers \\
        rad & Index of accessibility to radial highways per town \\
        tax & Property-tax rate per \$US10,000 per town \\
        ptr & Pupil–teacher ratio per town \\
        black & Percentage of blacks \\
      \B  lsp & Percentage of lower status population \\ \hline
    \end{tabular}
\end{table}

\section{Empirical Application - Boston Housing Dataset}\label{sec:app1:bh}

We now compare the ESF selection procedures using the Boston Housing Dataset, which was first used by \citet{hr78_bh} to evaluate the relationship between house prices and demand for clean air. \citet{gp96_sbh} later revisited the dataset when they noted the high spatial correlation in the dataset and proposed estimating a spatial error model instead. However, as there is no guarantee theirs is the correct specification, and given that the researcher is only concerned with the direct effect, ESF is an appropriate methodology allowing to simply control for the spatial effects.


The dataset includes 508 census tracts (spatial units). Table \ref{discr} describes the variables used in the analysis. The eigenvectors are from a binary SWM where the tracts are connected if they share a border, and SWM is normalised by the maximal of the row (or column) sum. The following basic model (excluding the eigenvectors) is:
\begin{align*}
    \ln(p_i) = &\beta_0 + \beta_1crim_i + \beta_2 zn_i +\beta_3 indus_i +\beta_4chas_i +\beta_5 nox^2_i+\beta_6rm_i+ \beta_7 age_i  \\
    & + \beta_8 dis_i + \beta_9 rad_i + \beta_{10}tax_i +\beta_{11} ptr_i + \beta_{12}black_i + \beta_{13}lsp_i +  \varepsilon_i.
\end{align*}

\begin{table}[ht!]
  \centering
  \caption{Parameter Estimation Results}
  \label{app_par}
\resizebox{0.9\textwidth}{!}{\begin{tabular}{@{\extracolsep{5pt}}lD{.}{.}{-3} D{.}{.}{-3} D{.}{.}{-3} D{.}{.}{-3} }
\hline
\hline
\T\B & \multicolumn{4}{c}{\textit{Dependent variable:} $\ln(p)$} \\
\cline{2-5}
\T & \multicolumn{1}{c}{simple-OLS} & \multicolumn{1}{c}{FstepZ} & \multicolumn{1}{c}{CV-pLasso} & \multicolumn{1}{c}{Mi-pLasso}\\
\B & \multicolumn{1}{c}{(1)} & \multicolumn{1}{c}{(2)} & \multicolumn{1}{c}{(3)} & \multicolumn{1}{c}{(4)}\\
\hline
\T crim & -0.010^{***} & -0.009^{***} & -0.010^{***} & -0.011^{***} \\
  & (0.002) & (0.002) & (0.002) & (0.001) \\
 zn & 0.001^{***} & 0.001^{*} & 0.001^{***} & 0.0003 \\
  & (0.0004) & (0.0004) & (0.0004) & (0.0002) \\
 indus & 0.002 & -0.0003 & 0.003 & 0.005^{***} \\
  & (0.002) & (0.002) & (0.002) & (0.001) \\
 chas & 0.104^{***} & 0.038 & 0.065^{***} & 0.053^{***} \\
  & (0.038) & (0.038) & (0.025) & (0.014) \\
 nox$^2$ & -0.588^{***} & -0.219^{*} & -0.126 & -0.316^{***} \\
  & (0.124) & (0.125) & (0.096) & (0.051) \\
 rm & 0.091^{***} & 0.177^{***} & 0.221^{***} & 0.169^{***} \\
  & (0.028) & (0.032) & (0.016) & (0.008) \\
 age & 0.0001 & -0.001^{*} & -0.001^{**} & -0.001^{***} \\
  & (0.001) & (0.001) & (0.0005) & (0.0002) \\
 dis & -0.047^{***} & -0.032^{***} & -0.029^{***} & -0.033^{***} \\
  & (0.008) & (0.007) & (0.006) & (0.004) \\
 rad & 0.014^{***} & 0.011^{***} & 0.011^{***} & 0.013^{***} \\
  & (0.003) & (0.003) & (0.002) & (0.001) \\
 tax & -0.001^{***} & -0.0004^{***} & -0.001^{***} & -0.001^{***} \\
  & (0.0001) & (0.0001) & (0.0001) & (0.0001) \\
 ptr & -0.039^{***} & -0.006 & -0.018^{***} & -0.036^{***} \\
  & (0.004) & (0.005) & (0.004) & (0.003) \\
 black & -0.003^{***} & -0.005^{***} & -0.005^{***} & -0.005^{***} \\
  & (0.001) & (0.001) & (0.001) & (0.0004) \\
 lsp & -0.029^{***} & -0.020^{***} & -0.017^{***} & -0.022^{***} \\
  & (0.004) & (0.003) & (0.002) & (0.001) \\
\B  Const. & 4.031^{***} & 2.655^{***} & 2.554^{***} & 3.364^{***} \\
  & (0.243) & (0.277) & (0.157) & (0.086) \\
 \hline
\T Adj. R$^{2}$ & \multicolumn{1}{c}{0.785} & \multicolumn{1}{c}{0.896} & \multicolumn{1}{c}{0.901} & \multicolumn{1}{c}{0.978} \\
 Resid. S.E. & \multicolumn{1}{c}{0.189} & \multicolumn{1}{c}{0.132} & \multicolumn{1}{c}{0.129} & \multicolumn{1}{c}{0.061} \\
\B d.f. & \multicolumn{1}{c}{492} & \multicolumn{1}{c}{431} & \multicolumn{1}{c}{449} & \multicolumn{1}{c}{295} \\
 \hline
\multicolumn{5}{l}{\T \textit{Note:} $^{*}$p$<$0.1; $^{**}$p$<$0.05; $^{***}$p$<$0.01. Robust standard errors in parenthesis. } \\
\end{tabular}}
\end{table}

Table \ref{app_par} shows the parameter estimates (excluding eigenvectors) for OLS, which ignores the spatial correlation, Mi-pLasso, CV-pLasso, and FstepZ. These results show that some of the OLS estimates are biased by spatial dependence. For example, age had a positive (but insignificant) coefficient when the spatial dependence is ignored, but in the filtered estimates, the coefficient is negative and significant as expected; the coefficient on $nox^2$, $dis$, and $rm$ also have a downward bias. Additionally, the filtered estimates also give a substantially better fit of house prices, with Mi-pLasso having an adjusted R$^{2}$ of 0.978, implying an almost perfect fit of the data. Mi-pLasso standard errors are generally the same or smaller than the other estimator.

\begin{table}[t]
    \centering
  \caption{Computational time and Selected Eigenvectors}
     \label{nosel}
    \begin{tabular}{l c c c }
    \hline
\T \B     & FStepZ & CV-pLasso & Mi-pLasso \\ \hline
\T    Computational time (seconds) & 14.37  &  10.23 & 0.15 \\
       Number of Eigenvectors & \multicolumn{1}{c}{61} & \multicolumn{1}{c}{43} & \multicolumn{1}{c}{197}    \\
        Significant at 0.1\% level & 16 & 18 &  85 \\
        Significant at 1\% level & 7 & 7 & 40 \\
        Significant at 5\% level & 13 & 7 &  57 \\
        Significant at 10\% level & 8 & 3 & 14  \\
\B        Not significant & 17  & 8 &  1 \\
         \hline
           \multicolumn{4}{l}{\begin{minipage}{12cm}%
            \vspace{0.1cm}
       \footnotesize{\textit{Note:} computational times exclude spectral decomposition. }
     \end{minipage}%
     }
    \end{tabular}
\end{table}

Table \ref{nosel} shows the computational times, the number of selected eigenvectors, and their significance levels, for the three ESF estimators. There is substantial variation in the number of selected eigenvectors between the procedures. Mi-Lasso selected over four and three times more eigenvectors than CV-Lasso and FstepZ. However, despite selecting substantially more eigenvectors for Mi-Lasso, only 0.5 percent of selected eigenvectors are insignificant compared to 28 percent and 21 percent for FstepZ and CV-Lasso. Mi-Lasso has more eigenvectors with coefficients significant at the 0.1 percent level than FstepZ or CV-Lasso selected in total, implying these techniques may be under-selecting in this case. Mi-Lasso is also over 65 times faster than both FstepZ and CV-Lasso.\footnote{The code to replicate the results in the section can be found in the attached filed `boston\_comp.R'.}

\section{Conclusion and Further Work}\label{conc1}

In this paper we have formalised the ESF assumptions and evaluated the existing solutions to the ESF eigenvector selection problem. Our analysis of existing procedures has shown that a dominant selection procedure currently does not exist. The forward-iterative procedures with a user-defined cut-off and eigenvector inclusion criterion can be viewed as ad hoc and are slow, especially as the sample size increases. \citep{seya15} proposed using Lasso with prediction accuracy CV to estimate the tuning parameter. However, as ESF aims to reduce bias on $\bm{\beta}_0$ rather than improve prediction accuracy, it is unclear if this is the best way to estimate the tuning parameter. Additionally, CV-based Lasso procedure is also slow, especially when $n$ is large.

We have proposed an alternative Lasso-based procedure called Morans’ $I$ Lasso (Mi-Lasso) that uses information about the level of spatial correlation in the naïve regression residuals to determine a point estimate for the Lasso tuning parameter instead of using CV. The key benefits of Mi-Lasso are that it is intuitive, theoretically grounded, and substantially faster than \citep{seya15} CV Lasso or stepwise procedures and can thus be implemented on large data sets. We have derived performance bounds for the Mi-Lasso estimates of the eigenvectors coefficients and shown the conditions necessary for the estimator to provide consistent eigenvector selection. Our simulation results confirm the estimator performs well in terms of bias and MSE compared to existing selection procedures for a range of levels of spatial correlation and in an empirical application on house prices. Additionally, we have shown using a property of the spectral decomposition and a simulations experiment, that ESF is robust to the presence of an unknown number of higher-order spatial lags in underlying DGP.

A key limitation of the ESF literature is that there are no results on constructing robust standard errors. As all the proposed procedures can be viewed as post-model selection estimators. Thus, all the corresponding estimators suffer from the corresponding post-model selection inference problem \citep{lp08}. Given the spatial dependence in the model, debiasing techniques such as Double Lasso \citep{bch14} or Partial Lasso \citep{CHS15} will not work well. A promising avenue of future research in the ESF literature is to extend Mi-Lasso (and other procedures), so standard errors robust to selection mistakes and the spatial dependence in the model can be calculated.

\bigskip

\noindent\textbf{Conflict of Interest Statement:} the authors declare no conflicts of interest

\bibliographystyle{agsm}
\bibliography{reference} 
 
\begin{appendices} 
  \section{Proof of theorems}\label{app:proofs}

  \begin{proof}[Proof of Lemma~\ref{lem:hdfwl}]
    Two important points to note is by Assumption \ref{as:spatial}.2 the $n\times k$ matrix $\bm{X}$ has full column rank and only the coefficient of the matrix $\bm{E}$ are being penalized. The objective function in \eqref{Lassofull} is coercive (for minimization) and strictly convex, thus, $[\hat{\bm{\beta}},\hat{\bm{\gamma}}]$ is a unique global minimizer. \eqref{Lassofull} is also subdifferentiable, specifically from the Karush-Kuhn-Tucker conditions for Lasso we have:
    \begin{align}
      \bm{X}'(\bm{y}-\bm{X}\hat{\bm{\beta}}-\bm{E}\hat{\bm{\gamma}})&=0 \label{kkt1}\\
      \bm{E}'(\bm{y}-\bm{X}\hat{\bm{\beta}}-\bm{E}\hat{\bm{\gamma}})-\frac{1}{Z^2}s(\hat{\bm{\gamma}})&=0 \label{kkt2}
  \end{align}
  where $s(\cdot)$ maps a positive entry to 1, a negative entry to -1 and zero to $\in[-1,1]$. Rearranging \eqref{kkt1} to make $\hat{\bm{\beta}}_\frac{1}{Z}$ the subject and substituting this into \eqref{kkt2} yields:
    \begin{align}
      \bm{E}'(\bm{y}-\bm{X}(\bm{X}'\bm{X})^{-1}\bm{X}'(\bm{y}-\bm{E}\hat{\bm{\gamma}})-\bm{E}\hat{\bm{\gamma}})-\frac{1}{Z^2}s(\hat{\bm{\gamma}})&=0 \nonumber \\
        \bm{E}'((\bm{I}-\bm{X}(\bm{X}'\bm{X})^{-1}\bm{X}')\bm{y}-(\bm{I}-\bm{X}(\bm{X}'\bm{X})^{-1}\bm{X}')\bm{E}\hat{\bm{\gamma}})-\frac{1}{Z^2}s(\hat{\bm{\gamma}})&=0 \nonumber \\
        \bm{E}'(\tilde{\bm{y}}-\tilde{\bm{E}}\hat{\bm{\gamma}})-\frac{1}{Z^2}s(\hat{\bm{\gamma}})&=0 \label{fwl}
    \end{align}

  The left-hand side of \eqref{fwl} is a sub-vector of the objective function in \eqref{Lassofwl} at $\bm{\gamma}=\hat{\bm{\gamma}}$ and it equals 0, thus, $\hat{\bm{\gamma}}=\tilde{\bm{\gamma}}$, if the minimisation is unique.
  \end{proof}

  \begin{proof}[Proof of Theorem~\ref{the:parcon}]
    By definition, $
      \tilde{\bm{\gamma}}=\arg \min_{\bm{\gamma}}||\tilde{\bm{y}}-\tilde{\bm{E}}\bm{\gamma}||^2_2+\frac{1}{Z^2}||\bm{\gamma}||_1$.
    Denoting $\bm{\Delta}=\tilde{\bm{\gamma}}-\bm{\gamma}_0$, then by the optimality of $\tilde{\bm{\gamma}}$ and dividing by $n$ we obtain:
    \begin{align}
      ||\tilde{\bm{y}}-\tilde{\bm{E}}\tilde{\bm{\gamma}}||^2_2/n +\frac{1}{Z^2n}||\tilde{\bm{\gamma}}||_1 & \enspace\leq\enspace
      ||\tilde{\bm{y}}-\tilde{\bm{E}}\bm{\gamma}_0||^2_2/n +\frac{1}{Z^2n}||\bm{\gamma}_0||_1 \nonumber \\
      ||\tilde{\bm{y}}-\tilde{\bm{E}}\tilde{\bm{\gamma}}||^2_2/n - ||\tilde{\bm{y}}-\tilde{\bm{E}}\bm{\gamma}_0||^2_2/n&\enspace\leq \enspace  \frac{1}{Z^2n}(||\bm{\gamma}_0||_1-||\tilde{\bm{\gamma}}||_1) \label{basicin2}
    \end{align}
Given $\bm{\Delta}_{\grave{\Omega}}=\tilde{\bm{\gamma}}_{\grave{\Omega}}$, $\bm{\gamma}_0=\bm{\gamma}_{\Omega}$ and the reverse triangle inequality
    $||\tilde{\bm{\gamma}}_\Omega||_1\geq ||\bm{\gamma}_{\Omega}||_1-||\bm{\Delta}_{\Omega}||_1$, we have:
    \begin{align}
      ||\bm{\gamma}_0||_1-||\tilde{\bm{\gamma}}||_1 &= ||\bm{\gamma}_0||_1-(||\tilde{\bm{\gamma}}_\Omega||_1+||\tilde{\bm{\gamma}}_{\grave{\Omega}}||_1)= ||\bm{\gamma}_0||_1-(||\tilde{\bm{\gamma}}_\Omega||_1+||\bm{\Delta}_{\grave{\Omega}}||_1) \nonumber \\
      ||\bm{\gamma}_0||_1-||\tilde{\bm{\gamma}}||_1 &\leq ||\bm{\gamma}_0||_1 - (||\bm{\gamma}_{\Omega}||_1-||\bm{\Delta}_{\Omega}||_1+||\bm{\Delta}_{\grave{\Omega}}||_1)\leq ||\bm{\Delta}_{\Omega}||_1-||\bm{\Delta}_{\grave{\Omega}}||_1 \label{basicpen}
    \end{align}
Furthermore:
    \begin{align}
      ||\tilde{\bm{y}}-\tilde{\bm{E}}\tilde{\bm{\gamma}}||^2_2/n - ||\tilde{\bm{y}}-\tilde{\bm{E}}\bm{\gamma}_0||^2_2/n& =||\tilde{\bm{E}}(\tilde{\bm{\gamma}}-\bm{\gamma}_0)-\bm{v}||_2^2/n-||\bm{v}||_2^2/n  \nonumber \\
  & = ||\tilde{\bm{E}}\bm{\Delta}||_2^2/n-2\bm{v}'\tilde{\bm{E}}\bm{\Delta}/n \nonumber \\
  & \geq_{(i)} ||\tilde{\bm{E}}\bm{\Delta}||_2^2/n- 2||\bm{v}'\tilde{\bm{E}}||_\infty/n ||\bm{\Delta}||_1 \nonumber \\
  & \geq_{(ii)} ||\tilde{\bm{E}}\bm{\Delta}||_2^2/n- \frac{1}{Z^2bn} ||\bm{\Delta}||_1 \label{inevent}
  \end{align}
($i$) uses H\"{o}lder inequality with $\ell_\infty$ and $\ell_1$ norms, $2|\bm{v}'\tilde{\bm{E}}(\bm{\Delta})|/n\leq 2||\bm{v}'\tilde{\bm{E}}||_\infty/n ||\bm{\Delta}||_1$.
  ($ii$) uses the event
  \begin{equation}\label{eventT}
    T := b2||\bm{v}'\tilde{\bm{E}}||_\infty/n \leq \frac{1}{Z^2n}
  \end{equation}
   where $b\geq1$ is an arbitrary constant ensuring the penalty dominates the random process. Combining \eqref{basicin2}, \eqref{basicpen} and \eqref{inevent}:
  \begin{align}
      ||\tilde{\bm{E}}\bm{\Delta}||_2^2/n & \leq \frac{1}{bZ^2n} (||\bm{\Delta}_{\Omega}||_1+||\bm{\Delta}_{\grave{\Omega}}||_1) +\frac{1}{Z^2n}(||\bm{\Delta}_{\Omega}||_1-||\bm{\Delta}_{\grave{\Omega}}||_1) \label{re2} \\
        & \leq \bigg(1 + \frac{1}{b}\bigg)\frac{1}{Z^2n} ||\bm{\Delta}_{\Omega}||_1-\bigg(1 - \frac{1}{b}\bigg)\frac{1}{Z^2n}||\bm{\Delta}_{\grave{\Omega}}||_1 \label{re1}
  \end{align}

  Given $||\tilde{\bm{E}}\bm{\Delta}||_2^2/n > 0$ and using \eqref{re1} we have $||\bm{\Delta}_{\grave{\Omega}}||_1 \leq \bar{b}||\bm{\Delta}_{\Omega}||_1$,
  where $\bar{b}=(b+1)/(b-1)$, allowing us to use the restricted eigenvalue condition RE$(\bar{b})$. Substituting in for $||\tilde{\bm{E}}\bm{\Delta}||_2^2/n$ in \eqref{re2} gives:
  \begin{align}
      \tau_{min}^2||\bm{\Delta}||_2^2 &\leq \frac{1}{bZ^2n} ||\bm{\Delta}||_1 +\frac{1}{Z^2n}(||\bm{\Delta}_{\Omega}||_1-||\bm{\Delta}_{\grave{\Omega}}||_1) \nonumber \\
      &\leq \bigg(\frac{1}{b}+1\bigg) \frac{1}{Z^2n} ||\bm{\Delta}_{\Omega}||_1 \nonumber \\
        &\leq \bigg(\frac{1}{b}+1\bigg) \frac{\sqrt{s}}{Z^2n} ||\bm{\Delta}_{\Omega}||_2 \label{l12l2}
  \end{align}
where the last inequality uses $||\bm{\Delta}_{\Omega}||_1\leq\sqrt{s}||\bm{\Delta}_{\Omega}||_2$ which holds by the Cauchy-Schwarz inequality. This implies the following $\ell_2$ parameter bound, which is \eqref{parcon2}:
  \begin{equation*}
    ||\tilde{\bm{\gamma}}-\bm{\gamma}_0||_2 \leq \frac{\big(\frac{1}{b}+1\big)\sqrt{s} }{\tau_{min}^2Z^2n}
  \end{equation*}
Again we can swap the $\ell_2$ norm for the $\ell_1$ norm and rearrange to give \eqref{parcon1}:
  \begin{equation*}
    ||\tilde{\bm{\gamma}}-\bm{\gamma}_0||_1 \leq \frac{\big(\frac{1}{b}+1\big)s }{\tau_{min}^2Z^2n}
  \end{equation*}
Similarly substituting RE$(\bar{b})$ in for $||\bm{\Delta}_{\Omega}||_2$ in \eqref{re2} and given \eqref{l12l2} yields
  \begin{align*}
        ||\tilde{\bm{E}}\bm{\Delta}||_2^2/n &\leq \bigg(\frac{1}{b}+1\bigg) \frac{\sqrt{s}}{Z^2\tau n^{3/2}} ||\tilde{\bm{E}}\bm{\Delta}_{\Omega}||_2
  \end{align*}
This implies the following $\ell_2$ performance bound, which is \eqref{parcon3}.
  \begin{equation*}
    \frac{1}{\sqrt{n}}||\tilde{\bm{E}}(\tilde{\bm{\gamma}}-\bm{\gamma}_0)||_2 \leq \frac{\big(\frac{1}{b}+1\big)\sqrt{s} }{\tau_{min}Z^2n}
  \end{equation*}

We have obtained \eqref{parcon1}, \eqref{parcon2} and \eqref{parcon3} by assuming \eqref{eventT}, we now need to evaluate the probability it is true, i.e. $P(T)$. Let $\frac{1}{Z^2}=t$ and using the definition of $||\cdot||_\infty$ we can rewrite \eqref{eventT} as:
  \[
  T := \max_{j\in N}2b|\bm{v}'\tilde{\bm{e}}_j| \leq t
  \]
where $\tilde{\bm{e}}_j$ is the $jth$ column of $\tilde{\bm{E}}$. By a union bound
  \begin{equation}
   P(\grave{T}) = P (\max_{j\in N}2b|\bm{v}'\tilde{\bm{e}}_j| \geq t) \leq n\max_{j\in N}P (2b|\bm{v}'\tilde{\bm{e}}_j| \geq t)
  \end{equation}
Given $\bm{v}$ is sub-Gaussian $(0, \sigma_{\bm{v}} )$ and $\tilde{\bm{e}}_j$ is a vector of real numbers, $\bm{v}'\tilde{\bm{e}}_j$ is also sub-gaussian:
  \begin{align*}
      P(\grave{T})  \leq n\max_{j\in N}P (2b|\bm{v}'\tilde{\bm{e}}_j|/n \geq t) \leq 2n\exp\bigg(-\frac{n^2t^2}{2\sigma^2_{\bm{v}}\max_{j\in N}||\tilde{\bm{e}}_j||^2_2}\bigg) \leq 2n\exp\bigg(-\frac{nt^2}{2\sigma^2_{\bm{v}}}\bigg)
  \end{align*}
  where the final inequality holds by assuming $\max_{j\in N}||\tilde{\bm{e}}_j||_2\leq \sqrt{n}$. Letting $t=\sqrt{\frac{4\sigma^2_{\bm{v}}\log n}{n}}$ we get $P(\grave{T})  \leq \frac{2}{n}$, therefore $P(T) =1 -P(\grave{T})  \geq 1- \frac{2}{n}$, and $P(T) \to 1 $ as $n\to\infty$.
  \end{proof}

  To prove Proposition \ref{prop1} we state Lemma \ref{lem1}, which is a direct consequence of the Karush-Kuhn-Tucker conditions:
  \begin{lem}\label{lem1}
    $\tilde{\bm{\gamma}}= (\tilde{\gamma}_1,\ldots,\tilde{\gamma}_j,\ldots, \tilde{\gamma}_n)$ are the Lasso estimates defined by \eqref{Lassofwl} if and only if
    \begin{align*}
      \frac{d||\tilde{\bm{y}}-\tilde{\bm{E}}\gamma||^2_2}{d\gamma_j}|_{\gamma_j=\tilde{\gamma}_j}=&\frac{1}{Z^2}\sign(\tilde{\gamma_j})  & for \; j : \tilde{\gamma_j}\neq 0\\
      \bigg|\frac{d||\tilde{\bm{y}}-\tilde{\bm{E}}\gamma||^2_2}{d\gamma_j}\bigg|_{\gamma_j=\tilde{\gamma}_j} \leq &\frac{1}{Z^2}  & for \; j : \tilde{\gamma}_j= 0
   \end{align*}
  \end{lem}

  \begin{proof}[Proof of Propostition~\ref{prop1}]
    By definition:
    $$\tilde{\bm{\gamma}}= \arg\min_{\bm{\gamma}} [(\tilde{\bm{y}}-\tilde{\bm{E}}\bm{\gamma})'(\tilde{\bm{y}}-\tilde{\bm{E}}\bm{\gamma})+\frac{1}{Z^2}||\bm{\gamma}||_1]$$
    Let $\bm{\Delta}=\tilde{\bm{\gamma}}-\bm{\gamma}_0$ and define
    $$\bm{d}(\bm{\Delta})= \big[(\tilde{\bm{y}}-\tilde{\bm{E}}(\bm{\gamma}+\bm{\Delta}))'(\tilde{\bm{y}}-\tilde{\bm{E}}(\bm{\gamma}+\bm{\Delta}))\big]-(\tilde{\bm{y}} -\tilde{\bm{E}}\bm{\gamma})'(\tilde{\bm{y}}-\tilde{\bm{E}}\bm{\gamma})  + \frac{1}{Z^2}||\bm{\gamma}+\bm{\Delta}||_1 $$
    Then
  \begin{align}
    \bm{\Delta} = \arg\min_{\bm{\Delta}} \bm{d}(\bm{\Delta}) \label{u}
  \end{align}
  Splitting $\bm{d}(\bm{\Delta})$ into two parts $\bm{d}_1(\bm{\Delta})$ and $\bm{d}_2(\bm{\Delta})$. Let
  \begin{align*}
    \bm{d}_1(\bm{\Delta}) & = \big[(\tilde{\bm{y}}-\tilde{\bm{E}}(\bm{\gamma}+\bm{\Delta}))'(\tilde{\bm{y}}-\tilde{\bm{E}}(\bm{\gamma}+\bm{\Delta}))\big]-(\tilde{\bm{y}} -\tilde{\bm{E}}\bm{\gamma})'(\tilde{\bm{y}}-\tilde{\bm{E}}\bm{\gamma}) \\
    & = [(\tilde{\bm{v}}-\tilde{\bm{E}}\bm{\Delta})'(\tilde{\bm{v}}-\tilde{\bm{E}}\bm{\Delta})-\tilde{\bm{v}}'\tilde{\bm{v}}] \\
    & = - 2\bm{\Delta}'\tilde{\bm{E}}'\tilde{\bm{v}}  +\bm{\Delta}'\tilde{\bm{E}}'\tilde{\bm{E}}\bm{\Delta} \\
    & = -2(\sqrt{n}\bm{\Delta})'\bm{z}+(\sqrt{n}\bm{\Delta})'\bm{C}(\sqrt{n}\bm{\Delta})
  \end{align*}
  where $\bm{z}=\tilde{\bm{E}}'\tilde{\bm{v}}/\sqrt{n}$. Differentiate $\bm{d}_1(\bm{\Delta})$ w.r.t. $\bm{\Delta}$

  \begin{equation}\label{diff}
    \frac{d\bm{d}_1(\bm{\Delta})}{d\bm{\Delta}} = 2\sqrt{n}\big(\bm{C}(\sqrt{n}\bm{\Delta})-\bm{z}\big)
  \end{equation}

  Now assuming that $\bm{\Delta}$ exists such that $\bm{\Delta}_{\grave{\Omega}} =0$ and  $\bm{\Delta}_{\Omega}$ is the solution of:
  \begin{align}
      \bm{C}_{\Omega\Omega}(\sqrt{n}\bm{\Delta}_{\Omega})-\bm{z}_{\Omega}=-\frac{1}{2Z^2\sqrt{n}}\sign(\bm{\gamma}_{\Omega}) \label{h1}
  \end{align}

  Event A ensures:
  \begin{equation}\label{eq2}
    |(\bm{C}_{\Omega\Omega})^{-1}\bm{z}_{\Omega}|<\sqrt{n}(|\bm{\gamma}_{\Omega}|-\frac{1}{2Z^2n}|(\bm{C}_{\Omega\Omega})^{-1}\sign(\bm{\gamma}_{\Omega})|)
  \end{equation}

  Event B and the IC ensure:
  \begin{equation}\label{eq3}
    |\bm{C}_{\grave{\Omega}\Omega}(\bm{C}_{\Omega\Omega})^{-1}\bm{z}_{\Omega}-\bm{z}_{\grave{\Omega}}|\leq \frac{1}{2Z^2\sqrt{n}}(\bm{1}-|\bm{C}_{\grave{\Omega}\Omega}(\bm{C}_{\Omega\Omega})^{-1}\sign(\bm{\gamma}_{\Omega})|)
  \end{equation}

  Equations \eqref{h1} and \eqref{eq2} together imply $|\bm{\Delta}_{\Omega}|<|\bm{\gamma}_{\Omega}| \label{h2}$. Similarly, \eqref{h1} and \eqref{eq3} imply $-\frac{1}{2Z^2\sqrt{n}}\bm{1}\leq \bm{C}_{\grave{\Omega}\Omega}(\sqrt{n}\bm{\Delta}_{\Omega})-\bm{z}_{\grave{\Omega}}\leq \frac{1}{2Z^2\sqrt{n}}\bm{1}$. Thus, by Lemma \ref{lem1}, \eqref{u}, \eqref{diff} and the uniqueness of the Lasso solution, $\sign(\tilde{\bm{\gamma}}_{\Omega})= \sign(\bm{\gamma}_{\Omega})$ and $\tilde{\bm{\gamma}}_{\grave{\Omega}}= \bm{\Delta}_{\grave{\Omega}}=0$.
  \end{proof}

  \begin{proof}[Proof of Theorem~\ref{the:selcon}]
    This proof works by bounding the tail probability of Proposition \ref{prop1} using conditions on the disturbance term. By Proposition \ref{prop1} we have $\p\big(\tilde{\bm{\gamma}}=_s\bm{\gamma}_0\big)\geq \p(A \cap B)$, thus:
    \begin{align}
      1-\p(A \cap B) &\leq \p(\grave{A})+\p(\grave{B}) \nonumber\\
      & \leq \sum^s_{i=1}\p\bigg(|k_{i}|\geq\sqrt{n}\bigg(|\gamma_{i}|-\frac{1}{2Z^2n}b_{i}\bigg)\bigg)+ \sum^q_{i=1}\p\bigg(| \xi_{i}|\geq\frac{1}{2Z^2\sqrt{n}}\nu_{i}\bigg) \label{summ}
    \end{align}
  where $\bm{k}=(k_{1},\ldots,k_{s})'=\bm{C}_{\Omega\Omega}^{-1}\bm{z}_{\Omega}$, $\bm{\xi}=(\xi_{1},\ldots,\xi_{q})'=\bm{C}_{\grave{\Omega}\Omega}\bm{C}_{\Omega\Omega}^{-1}\bm{z}_{\Omega}-\bm{z}_{\grave{\Omega}}$
  and $\bm{b}=(b_{1},\ldots,b_{s})'=\bm{C}_{\Omega\Omega}^{-1}\sign(\bm{\gamma}_{\Omega})$. Now if we write $\bm{k}=\bm{H}_A'\bm{v}$ where $\bm{H}_A' =(\bm{h}_{1,a},\ldots,\bm{h}_{s,a})'=\bm{C}_{\Omega\Omega}^{-1}(n^{-\frac{1}{2}}\tilde{\bm{E}}_{\Omega})$, then:
  \[
  \bm{H}'_A\bm{H}_A=\bm{C}_{\Omega\Omega}^{-1}n^{-1}\tilde{\bm{E}}_{\Omega}'\tilde{\bm{E}}_{\Omega} \bm{C}_{\Omega\Omega}^{-1} = \bm{C}_{\Omega\Omega}^{-1}
  \]
  Therefore, using assumption \ref{as:hidem}.2 gives $k_{i}= \bm{h}_{i,a}'\tilde{\bm{v}}$ with
  \begin{equation}\label{m2}
    ||\bm{h}_{i,a}||^2_2\leq\frac{1}{M_2}\;\;\forall \; i=1,\dots,s.
  \end{equation}
  Similarly if we write $\bm{\xi}=\bm{H}_B'\tilde{\bm{v}}$ where $\bm{H}_B' =(\bm{h}_{1,b},\ldots,\bm{h}_{q,b})'=\bm{C}_{\grave{\Omega}\Omega}\bm{C}_{\Omega\Omega}^{-1}(n^{-\frac{1}{2}}\tilde{\bm{E}}_{\Omega}')-n^{-\frac{1}{2}}\tilde{\bm{E}}_{\grave{\Omega}}'$, then:
  \begin{align*}
    \bm{H}'_B\bm{H}_B &=(\bm{C}_{\grave{\Omega}\Omega}\bm{C}_{\Omega\Omega}^{-1}(n^{-\frac{1}{2}}\tilde{\bm{E}}_{\Omega}')-n^{-\frac{1}{2}}\tilde{\bm{E}}_{\grave{\Omega}}') (n^{-\frac{1}{2}}\tilde{\bm{E}}_{\Omega}\bm{C}_{\Omega\Omega}^{-1}\bm{C}_{\grave{\Omega}\Omega}-n^{-\frac{1}{2}}\tilde{\bm{E}}_{\grave{\Omega}}) \\
    &= n^{-1}\tilde{\bm{E}}_{\grave{\Omega}}'(I-\tilde{\bm{E}}_{\Omega}(\tilde{\bm{E}}_{\Omega}'\tilde{\bm{E}}_{\Omega})^{-1}\tilde{\bm{E}}_{\Omega}')\tilde{\bm{E}}_{\grave{\Omega}}
  \end{align*}
The eigenvalues of $I-\tilde{\bm{E}}_{\Omega}(\tilde{\bm{E}}_{\Omega}'\tilde{\bm{E}}_{\Omega})^{-1}\tilde{\bm{E}}_{\Omega}'$ are 0 and 1, therefore using assumption \ref{as:hidem}.1 we have $\xi_{i}= \bm{h}_{i,b}'\tilde{\bm{v}}$ with:
  \begin{equation}\label{m1}
    ||\bm{h}_{i,b}||^2_2\leq M_1\;\;\forall \; i=1,\dots,s.
  \end{equation}

  Also note that:
  \begin{equation}\label{bbound}
    \bigg|\frac{1}{Z^2n}\bm{b}\bigg|=\bigg|\frac{1}{Z^2n}\bm{C}_{\Omega\Omega}^{-1}\sign(\bm{\gamma}_{\Omega})\bigg|\leq \frac{1}{Z^2M_2n}||\sign(\bm{\gamma}_{\Omega})||_2= \frac{1}{Z^2M_2n}\sqrt{s}
  \end{equation}

  Given \eqref{m2}, \eqref{m1} and Assumption \ref{as:spatial}.4 $\E[v_i^{4}]<\infty$,  implies $\E[k_{i}^{4}]<\infty$ and $\E[\xi_{i}^{4}]<\infty$. In fact for any given constant n-dimensional vector $\bm{\alpha}$, $\E(\bm{\alpha}'\bm{v})^{4}\leq (3)!!||\bm{\alpha}||^2_2\E[v_i^{4}]$. The tail probability of an \textit{i.i.d.} random variable with bound $4^{th}$ moments is bounded by:
  \begin{equation}\label{tbound}
    \p(k_{i}>t)=O(t^{-4})
  \end{equation}
  Rearranging the first summation term of \eqref{summ}:
  \begin{equation*}
    \sum^s_{i=1}\p\bigg(|k_{i}|\geq\sqrt{n}\bigg(|\gamma_{i}|-\frac{b_{i}}{2Z^2n}\bigg)\bigg)= \sum^s_{i=1}\p\bigg(|k_{i}|\geq\frac{b_{i}}{2Z^2\sqrt{n}}\bigg(\frac{2nZ^2|\gamma_{i}|}{b_{i}}-1\bigg)\bigg)
  \end{equation*}
  Next, we can use \eqref{tbound} to bound the $s$ probabilities:
  \begin{equation}\label{qbound}
    \sum^s_{i=1}\p\bigg(|k_{i}|\geq\frac{b_{i}}{2Z^2\sqrt{n}}\bigg(\frac{2nZ^2|\gamma_{i}|}{b_{i}}-1\bigg)\bigg) =\sum^s_{i=1} O\Bigg(\bigg(\frac{b_{i}}{2Z^2\sqrt{n}}\bigg)^{-4}\bigg(\frac{2nZ^2|\gamma_{i}|}{b_{i}}-1\bigg)^{-4}\Bigg)
  \end{equation}
  We now need to evaluate the bounds for both terms in \eqref{qbound}. For the second term we use \eqref{bbound} and Assumption \ref{as:hidem}.4 to replace $\frac{nZ^2}{b_{i}}$ by $\frac{nZ^2M_2}{\sqrt{s}}$ and its associated $O(n^{c_1})$ bound:
  \[
  O\Bigg(\bigg(\frac{2nZ^2|\gamma_{i}|}{b_{i}}-1\bigg)^{-4}\bigg)=O\Bigg( \bigg(\frac{2nZ^2M_2|\gamma_{i}|}{\sqrt{s}}-1\bigg)^{-4}\Bigg)=
  O\Bigg(\bigg(\frac{2nZ^2M_2|\gamma_{i}|}{O(n^{\frac{c_1}{2}})}-1\bigg)^{-4}\Bigg)
  \]

  Assumption \ref{as:hidem}.3 bounds $\sqrt{n}|\gamma_{i}|$ and $\frac{1}{Z^2\sqrt{n}}= o_p(n^{\frac{c_1-c_2}{2}})$. The bounds containing powers of $n$ cancel out, the remaining term depends only $M_2$ and $M_3$ and is therefore $O(1)$:
  \[
  O\Bigg(\bigg(\frac{2nZ^2M_2|\gamma_{i}|}{O(n^{\frac{c_1}{2}})}-1\bigg)^{-4}\Bigg)=
  O\Bigg(\bigg(\frac{o_p(n^{\frac{c_1-c_2}{2}})2n^{\frac{c_2}{2}}M_3M_2}{O(n^{\frac{c_1}{2}})}-1\bigg)^{-4}\Bigg)=O_p(1)
  \]

 Note that Assumption \ref{as:hidem}.3 is used with equality $\forall \: |\gamma_{i}|$ rather than as an inequality on $\min |\gamma_{i}|$. This is because of the $-4$ exponent which implies that the highest bound will be obtained for the smallest value of the expression in brackets. Expression \eqref{qbound} now reduces to:
  \[
  \sum^s_{i=1} O\Bigg(\bigg(\frac{b_{i}}{2Z^2\sqrt{n}}\bigg)^{-4}\bigg(\frac{2nZ^2|\gamma_{i}|}{b_{i}}-1\bigg)^{-4}\Bigg)= s O\Bigg(\bigg(\frac{b_{i}}{2Z^2\sqrt{n}}\bigg)^{-4}\Bigg) = s O\Bigg(\frac{16Z^{8}n^2}{b_{i}^{4}}\Bigg)
  \]

  Using \eqref{bbound} again to replace $\frac{nZ^2}{b_i}$ by $\frac{nM_2Z^2}{\sqrt{s}}$, integrating $s$ into the bound and ignoring the constants:
  \[
  s O\bigg(\frac{16Z^{8}n^2}{b_{i}^{4}}\bigg) = sO\bigg(\frac{16Z^{8}n^2M_2^2}{s^2}\bigg) = O\bigg(\frac{Z^{8}n^2}{s}\bigg)
  \]

  Note that because $s+q=n$ and $s,n> 0$, it must be that:
  \begin{equation*}
  \frac{Z^{8}n^2}{s} < n^{3}Z^{8}
  \end{equation*}

  This is because the left hand side denominator is smaller by a factor $s$ and right hand side larger by a factor $n$. Therefore:
  \begin{equation*}
  O\bigg(\frac{Z^{8}n^2}{s}\bigg)=o(n^{3}Z^{8})
  \end{equation*}
  Thus,
  \begin{equation}\label{suma}
    \sum^s_{i=1}\p\bigg(|k_{i}|\geq\sqrt{n}\bigg(|\gamma_{i}|-\frac{b_{i}}{2Z^2n}\bigg)\bigg)= o(n^{3}Z^{8})
  \end{equation}

  For the second summation term of \eqref{summ}, using \eqref{tbound} we have:
  \begin{equation*}
  \sum^q_{i=1}\p\bigg(| \xi^n_i|\geq\frac{1}{2Z^2\sqrt{n}}\nu_i\bigg)  = \sum^q_{i=1}O\bigg(\bigg(\frac{1}{2Z^2\sqrt{n}}\nu_i\bigg)^{-4}\bigg)
  \end{equation*}

  As $\nu_i$ (by Assumption \ref{as:hidem}) and 2 are both constant they can be ignored
  \begin{equation*}
   \sum^q_{i=1}O\bigg(\bigg(\frac{1}{2Z^2\sqrt{n}}\nu_i\bigg)^{-4}\bigg) = qO\big(Z^{8}n^2\big)
  \end{equation*}

  Integrating $q$ into the bound and noting that $s<n$, so if $s=O(n)$ then $q=O(n)$
  \begin{equation*}
    qO\big(Z^{8}n^2\big)=O(n^3Z^{8})
  \end{equation*}

   We therefore have the following bound:
   \begin{equation}\label{sumb}
    \sum^q_{i=1}\p\bigg(| \xi^n_i|\geq\frac{1}{2Z^2\sqrt{n}}\nu_i\bigg) = O(n^3Z^{8})
   \end{equation}

   Which provides the following asymptotic lower bound on the probability of the intersection for $\frac{1}{n^3Z^{8}}\to \infty$:
   \begin{align}
     \p(A \cap B) \geq 1 - o(n^{3}Z^{8}) - O(n^3Z^{8}) \to 1 \;\;\; as \;n \to \infty
   \end{align}
  \end{proof}

  \newpage
  \section{Supplementary figures}\label{app:figures}

  \begin{figure}[bh!]
    \caption{Bias and MSE of $\beta$ and number of selected eigenvectors, setup A with $\mu=4$}
    \label{fig:mu4}
    \centering
      \includegraphics[width=0.9\textwidth]{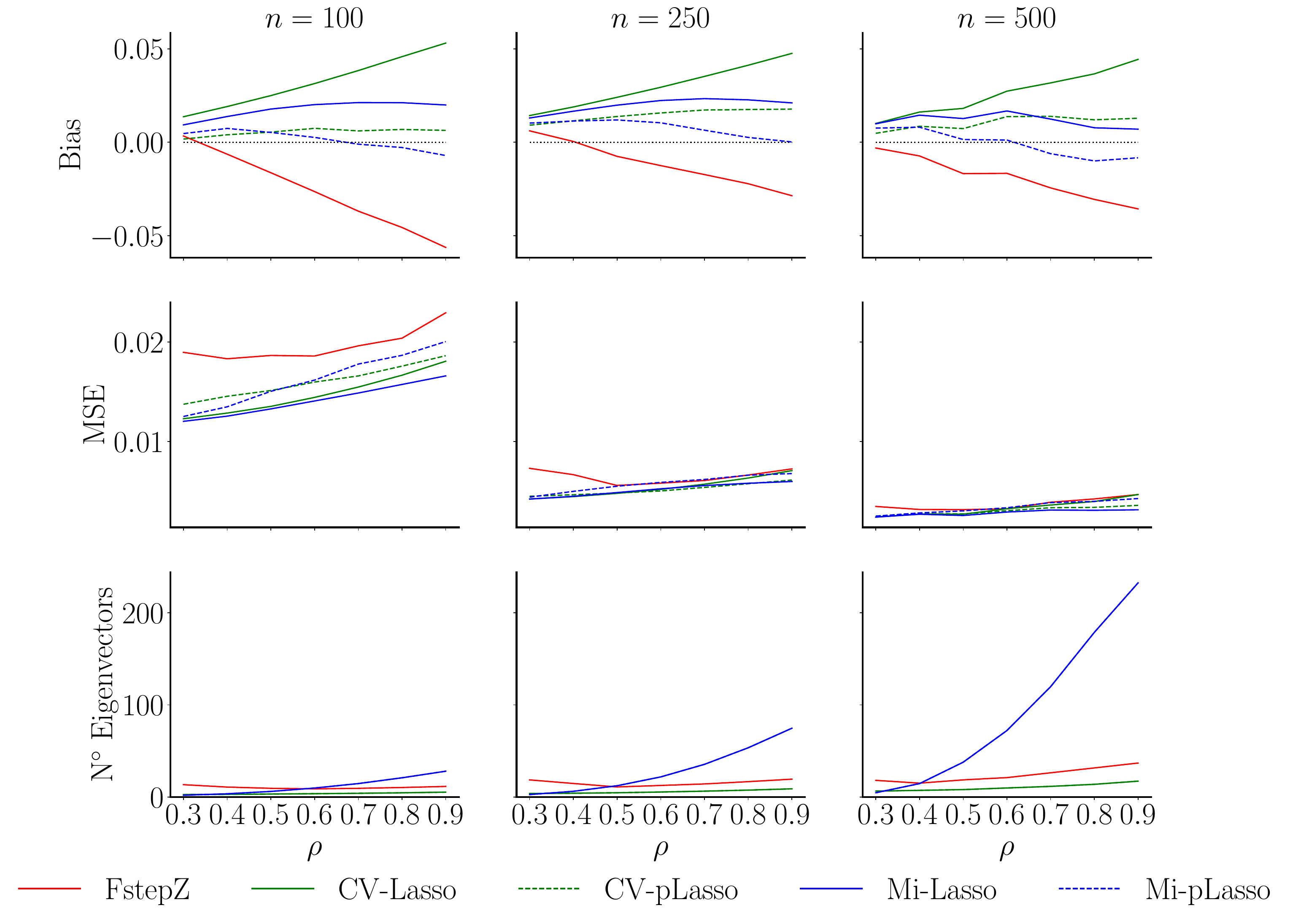}
    \bigskip
    \caption{Bias and MSE of $\beta$ and number of selected eigenvectors, setup A with $\mu=8$}
    \label{fig:mu8}
    \centering
      \includegraphics[width=0.9\textwidth]{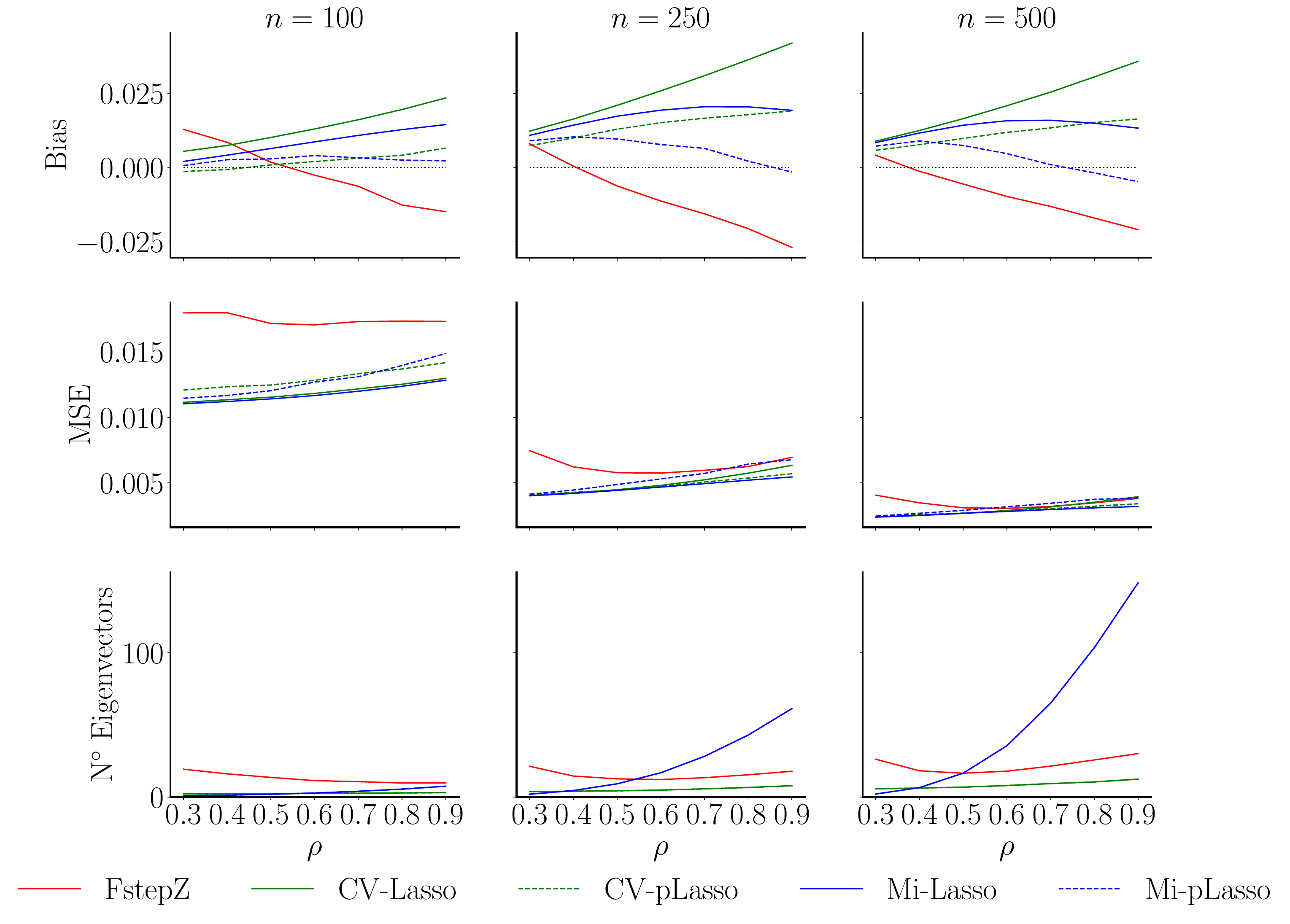}
  \end{figure}

\end{appendices}

\end{document}